\documentclass{aa}  
\usepackage{graphicx}
\usepackage{txfonts}
\usepackage{colortbl}

\usepackage[dvipsnames]{xcolor}
\usepackage{booktabs}
\usepackage{natbib,twoopt}
\usepackage[breaklinks=true,colorlinks=true,linkcolor=blue,citecolor=blue]{hyperref} 
\bibpunct{(}{)}{;}{a}{}{,}    
\usepackage{tabularx}
 
\usepackage{longtable}
\usepackage{lscape}
\usepackage{rotating}
\usepackage{ulem}

\colorlet{tableheadcolor}{white!25} 
\newcommand{\headcol}{\rowcolor{tableheadcolor}} %
\colorlet{tablerowcolor}{gray!10} 
\newcommand{\rowcol}{\rowcolor{tablerowcolor}} %

\let\ACMmaketitle=\maketitle
\renewcommand{\maketitle}{\begingroup\let\footnote=\thanks \ACMmaketitle\endgroup}

\begin{document}

   \title{VST-GAME: Galaxy assembly as a function of mass and environment with VST }
   \subtitle{Photometric assessment and density field of MACSJ0416 \footnote{\footnotesize Full version of Table \ref{table:firstrows} is only available in electronic form at the CDS via anonymous ftp to \url{cdsarc.cds.unistra.fr} (130.79.128.5) or via \url{https://cdsarc.cds.unistra.fr/cgi-bin/qcat?J/A+A/}}} 
   
   \author{
   N. Estrada\inst{\ref{unipd},\ref{oapd}} \and 
   A. Mercurio\inst{\ref{oac},\ref{unisa}} \and 
   B. Vulcani\inst{\ref{oapd}}  \and 
   G. Rodighiero\inst{\ref{unipd},\ref{oapd}} \and 
   M. Nonino\inst{\ref{oats}} \and 
   M.~Annunziatella\inst{\ref{madrid}}\and
   P.~Rosati\inst{\ref{unife},\ref{inafbo}}\and 
   C.~Grillo\inst{\ref{unimilano},\ref{iasf-milano}}\and
   G.~B.~Caminha\inst{\ref{maxplank1}}\and
   G.~Angora\inst{\ref{unife},\ref{oac}} \and 
   A.~Biviano\inst{\ref{oats},\ref{ifpu}} \and
   M.~Brescia\inst{\ref{oac},\ref{unina}} \and
   G.~De Lucia\inst{\ref{oats}} \and
   R.~Demarco\inst{\ref{udec}} \and
   M.~Girardi\inst{\ref{oats},\ref{utrieste}}\and
   R.~Gobat\inst{\ref{pucv}} \and
   B.C.~Lemaux\inst{\ref{geminin}, \ref{ucdavis}}
   }

   \institute{
   Dipartimento di Fisica e Astronomia "Galileo Galilei", Universit\`a degli studi di Padova, Vicolo dell'Osservatorio, 3, I-35122, Padova, Italy 
   \email{gilbertonicolas.estradamartinez@phd.unipd.it}
   \label{unipd} \and
   INAF - Osservatorio Astronomico di Padova, Vicolo dell'Osservatorio, 5, I-35122, Padova, Italy 
   \label{oapd} \and
   INAF - Osservatorio Astronomico di Capodimonte, Via Moiariello 16, 80131 Napoli, Italy   \email{amata.mercurio@inaf.it}
   \label{oac}  \and
   Dipartimento di Fisica “E.R. Caianiello”, Universit\`a Degli Studi di Salerno, Via Giovanni Paolo II, I–84084 Fisciano (SA), Italy
   \label{unisa}  \and
   INAF - Osservatorio Astronomico di Trieste, via Tiepolo 11, 34143, Trieste, Italy
   \label{oats} \and
   Dipartimento di Fisica dell'Universit\`a degli Studi di Trieste, Sezione di Astronomia, via Tiepolo 11, I-34143 Trieste, Italy
   \label{utrieste} \and
   Centro de Astrobiología, Instituto Nacional de Técnica Aeroespacial, Ctra de Torrejón a Ajalvir, km 4, 28850 Torrejón de Ardoz, Madrid, Spain
   \label{madrid} \and
   Dipartimento di Fisica e Scienze della Terra, Universit\`a di Ferrara, Via Saragat 1, 44122 Ferrara, Italy
   \label{unife}  \and
   INAF - OAS, Osservatorio di Astrofisica e Scienza dello Spazio di Bologna, via Gobetti 93/3, I-40129 Bologna, Italy
   \label{inafbo} \and
   Dipartimento di Fisica, Universit\`a  degli Studi di Milano, via Celoria 16, I-20133 Milano, Italy
   \label{unimilano} \and
   INAF - IASF Milano, via A. Corti 12, I-20133 Milano, Italy 
   \label{iasf-milano}\and
   Max-Planck-Institut f\"ur Astrophysik, Karl-Schwarzschild-Str. 1,
   D-85748 Garching, Germany
   \label{maxplank1}\and
   Institute for Fundamental Physics of the Universe (IFPU), via Beirut 2, 34151, Trieste, Italy
   \label{ifpu}\and
   Dipartimento di Fisica "Ettore Pancini", Universit\`a degli Studi di Napoli Federico II, Via Cinthia, I-80126 Napoli, Italy
   \label{unina} \and
   Departamento de Astronom\'ia, Facultad de Ciencias F\'isicas y Matem\'aticas,
Universidad de Concepci\'on, Concepci\'on, Chile
   \label{udec} \and
   Instituto de Física, Pontificia Universidad Católica de Valparaiso, Avda. Universidad 330, Placilla, Valparaíso, Chile
   \label{pucv} \and
   Gemini Observatory, NSF's NOIRLab, 670 N. A'ohoku Place, Hilo, Hawai'i, 96720, USA \label{geminin} \and
   Department of Physics and Astronomy, University of California, One Shields Avenue, Davis, CA 95616, USA
   \label{ucdavis}
}
   \date{Received ; accepted}
  \abstract
   {Observational studies have widely demonstrated that galaxy physical properties are strongly affected by the surrounding environment. 
   On one side, gas inflows provide galaxies with new fuel for star formation. On the other side, the high temperatures and densities of the medium are expected to induce quenching in the star formation. 
   Observations of large structures, in particular filaments at the cluster outskirts (r>2r$_{200}$), are currently limited to the low redshift Universe. Deep and wide photometric data, better if combined with spectroscopic redshifts, are required to explore several scenarios on galaxy evolution at intermediate redshift.}
   {We present a multi-band dataset for the cluster MACS J0416.1-2403 (z=0.397), observed in the context of the Galaxy Assembly as a function of Mass and Environment with the VLT Survey Telescope (VST-GAME) survey. 
   The project is aimed at gathering deep ($r$<24.4) and wide (approx. 20x20Mpc$^2$) observations at optical ($u$, $g$, $r$, $i$, VST) wavelengths for six massive galaxy clusters at 0.2<z<0.6, complemented with near-infrared data ($Y$, $J$, $Ks$, VISTA, ESO public survey GCAV). 
   The aim is to investigate galaxy evolution in a wide range of stellar masses and environmental conditions. This work describes the photometric analysis of the cluster and the definition of a density field, which will be a key ingredient for further studies on galaxy properties in the cluster outskirts.}
   {We extracted sources paying particular attention to recovering the faintest ones and simultaneously flagging point sources and sources with photometry affected by artifacts in the images. We combined all the extractions in a multiband catalog that is used to derive photometric redshifts through spectral energy distribution (SED) fitting.  We then defined cluster memberships up to 5r$_{200}$ from the cluster core and measure the density field, comparing galaxy properties in different environments.}
   {We find that the $g-r$ colors show bimodal behaviors in all the environments, but the peak of the distribution of red galaxies shifts toward redder colors with increasing density, and the fraction of galaxies in the blue cloud increases with decreasing density. We also found three overdense regions in the cluster outskirts at r$\sim$5r$_{200}$. Galaxies in these structures have mean densities and luminosities similar to those of the cluster core. The color of galaxies suggests the presence of evolved galaxy populations, an insight into pre-processing phenomena over these substructures. We release the multiband catalog, down to the completeness limit of $r<24.4$ mag.
   }
   {}
\maketitle

\section{Introduction}
\label{sec:intro}

In the last decades, the exploration of the external regions of galaxy clusters has become one of the new test beds for studying cosmology and cluster astrophysics. The characterization of these regions allows us to understand the physics of the intracluster medium (ICM) and the intergalactic medium (IGM), and to interpret  X-ray and Sunyaev–Zeldovich observations mainly focused on the cluster core. In the hierarchical structure formation model, galaxy clusters grow and evolve through a series of mergers and accretion from the surrounding large-scale structures in their outer envelope \citep{Kauffmann99,Benson01,Springel05,Delucia06}. The accretion flows leave characteristic marks, especially in the cluster outskirts, giving rise to caustics in the dark matter density profile \citep{Mansfield17,Diemer17}, internal bulk and turbulent gas motions \citep{Lau09,Vazza09, Battaglia_2012}, nonequilibrium electrons in the ICM \citep{RuddNagai09}, and an inhomogeneous gas density distribution \citep{NagaiLau11,Roncarelli13}. Nonetheless, the full accretion physics taking place in the cluster outskirts still needs to be explored \citep{dekel09,Danovich12,Welker20,Walker19}, mainly because there are few complete observations, which also happen to be challenging, in the cluster outskirts, or envelopes, and comparisons with simulations are needed in this regime.
Up to now, most of the studies have focused on the relatively dense central regions of galaxy clusters, the inner $\sim$10\% in terms of volume, representing only a portion of ICM.

Galaxies interacting with the hot ionized gas of the ICM can experience a plethora of physical mechanisms. 
When they pass through the ICM at high velocities, they can suffer ram pressure stripping, which removes an important fraction of their cold gas \citep{gunngott72}. The ICM is also able to remove the galaxy’s warm gas through a mechanism known as starvation \citep{Larson80,McCarthy08}. 
Other mechanisms that galaxies can experience in their passage through the cluster core are tidal stripping \citep[e.g.,][]{Zwicky51,Gnedin03,Villalobos14}, thermal evaporation \citep[e.g.,][]{CowieSongalia77}, and galaxy-galaxy interactions or harassment \citep[e.g.,][]{spitzerbaade1951,Moore96,Moore99}. Typically, any form of gas removal or consumption entails a direct decrease or inhibition of future star formation, leading to galaxy quenching. As a consequence, galaxies in clusters have properties that differ from those in the field \citep{Dressler1980,Poggianti1999,Bai2009}.

As the properties of the ICM vary from the cluster core to the outskirts \citep{Nagai2011,Ichikawa2013,Lau2015,Biffi2018,Mirakhor2021}, galaxies in the external regions are expected to have a different degree of interaction with the ICM, and the efficiency of the aforementioned physical processes could be different. In the cluster envelopes, environmental effects could also accelerate the consumption or remove the gas reservoir before galaxies enter a cluster, a process known as pre-processing \citep[e.g.,][]{Zabludoff_and_Mulchaey1998,Mihos04,Fujita04}. Hence the cluster envelopes are a key region that infalling galaxies cross even before they reach the cluster core. To fully characterize the effect of the cluster environment on galaxies, it is mandatory to have a thorough characterization of the population of galaxies in the outskirts of clusters. An additional motivation to study this region at intermediate and high redshift is because pre-processing mechanisms do not work in the same way in the local Universe as they worked in the past \citep{vanderburg2020}.

At redshift $z<0.3$, observational studies have found that the properties of galaxies such as star formation, gas content, and color are indeed affected by the cluster environment at large clustercentric distances, up to $\sim$3 virial radii at most \citep[e.g.,][]{Solanes02,Lewis02,gomez03,Verdugo08,Park09,Braglia09,vonderlinden10,Dressler13,Haines15,Rhee17,Paccagnella16}. In particular, spiral galaxies with low star formation rates are found in the outskirts of clusters \citep{Couch98,Dressler99} and the fraction of blue and star-forming galaxies in the outskirts is intermediate between the field and core values \citep{Wetzel13,Haines15,Guglielmo18,Bianconi18,Just2019}. 

The studies mentioned above have analyzed the infall region considering all possible directions, while other works consider the fall through filaments only. 
Galaxies are indeed preferentially accreted into clusters through filaments \citep[e.g.,][]{Colberg99,Ebeling04,Castignani2022}, and to a lesser extent from other directions. 
Comparing the properties of galaxies falling into clusters along filaments or from other directions (isotropic infalling), there is an enhancement of quenching for galaxies in filaments both at z$<$0.15 \citep{martinez16,Salerno20} and at $0.43 < z < 0.89$ \citep{Salerno19}.

Until now, a few studies have focused on cluster outskirts up to a large distance from the core with deep data at intermediate redshift \citep{Lubin2009,Schirmer11,Lemaux2012,Verdugo12,lu12,Just2019,Sarron19,Lemaux2019,Tomczak2019}. The emerging picture is that cluster outskirts play a major role in cluster evolution as early as $z\sim 1.4$ \citep{Lemaux2019,vanderburg2020}. However, a deep systematic mapping of the cluster envelopes out to very large clustercentric distances, r$\sim$5r$_{200}$, is still missing beyond the local Universe. In this context, the goal of this paper is to study the galaxy cluster MACS J0416.1-2403, analyzing a large galaxy sample drawn from the cluster center out to the outskirts, at epochs when the galaxy population is still rapidly evolving (\citealt{pog06,des07}). 

MACS J0416.1-2403 (hereafter M0416; \citealt{ebe01}) is a massive (M=0.88$\pm$0.13 $\times$ 10$^{15} $M$_\odot$) X-ray luminous (L$_X \sim 10^{45}$ erg s$^{-1}$) cluster at redshift $z=0.397$ (\citealt{bal16}), and one of the strongest lenses on the sky \citep{ber22,mes22,van21,van19}.  M0416 was first observed by Hubble Space Telescope (HST) as part of the Cluster Lensing And Supernova survey with Hubble (CLASH) survey \citep{pos12}. Then the cluster was observed as part of the Hubble Frontier Fields (HFF) initiative \citep{lot17}, obtaining deep images (5$\sigma$ point-source detection limit of $\sim$29 AB-mag). M0416 is also a target of future James Webb Space Telescope (JWST) observations by the CAnadian NIRISS Unbiased Cluster Survey (CANUCS) \citep{Willott2022}. 

M0416 was also observed as part of the European Southern Observatory (ESO) Large Programme "Dark Matter Mass Distributions of Hubble Treasury Clusters and the Foundations of $\Lambda$CDM Structure Formation Models" (CLASH-VLT; \citealt{Rosati14}) with the VIsible Multi-Object Spectrograph (VIMOS) at the ESO Very Large Telescope (VLT). Using these data, \cite{bal16} confirmed an overall complex dynamical state of this cluster. Early works identified M0416 as a merger, given its unrelaxed X-ray morphology and the separation ($\sim$ 200 kpc in projection) of the two brightest cluster galaxies (BCGs; \citealt{mann2012}). \cite{bal16} showed the presence of two main subclusters and supported the hypothesis that they are being observed in a precollisional phase, in agreement with the findings from radio and deep X-ray data of \cite{ogr15}. Finally, \citet{OlaveRojas2018}, using photometric and spectroscopic data, dynamically determined the presence of substructures up to r<2r$_{200}$ around M0416.

Using CLASH HST data, \cite{zit13} performed the first strong lensing analysis, showing an elongated projected mass distribution in the cluster core, typical of merging clusters. Then, taking advantage of the CLASH and CLASH-VLT spectroscopic follow-up program, a subsequent strong lensing model was presented in \citet{Grillo15} and in \citet{Hoa16}, including spectroscopic data from the Grism Lens-Amplified Survey from Space \citep[GLASS,][]{Treu15}. Combining weak and strong lensing analyses tested the impact of line-of-sight mass structures \citep{Jau14, Ric14, jau15, Hoa16, chi18}. \cite{jau15} detected two main central mass concentrations, and two possible secondary ones to the SW and NE, both at $\sim$ 2$^{\arcmin}$ from the cluster center. A high-precision strong lens model was obtained exploiting MUSE observations of the cluster, which led to the identification of a large sample of spectroscopic multiple images \citep{Caminha17}. This model was improved using HFF images and including the mass component associated with the hot gas \citep{bon17,Bon18}, and the kinematic measurements of a large sample of clusters galaxies \citep{Ber19}. A state-of-the-art strong lensing model of M0416 is presented in \citet{ber22} and was obtained utilizing 237 spectroscopically confirmed multiple images, which is the largest sample of secure multiply lensed sources utilized to date, including also stellar kinematics information of 64 cluster galaxies and the hot-gas mass distribution of the cluster determined from Chandra X-ray observations. 

\cite{bon17}, from a combined analysis of X-ray and gravitational lensing, measured a projected gas-to-total mass fraction of approximately 10\% at 350 kpc from the cluster center and showed that the dark matter over total mass fraction is almost constant, out to more than 350 kpc. Moreover, \cite{annunziatella2017} showed that there is no significant offset between the cluster stellar and dark-matter components in the core of the cluster. 

This paper aims to present a complete catalog of the wide field photometry centered on M0416 using the VLT Survey Telescope (VST) in four optical bands and a first analysis of the galaxy populations across several density environments up to $\sim$5r$_{200}$. In Sect. \ref{sec:data}, we present a complete description of the VST-GAME survey together with a brief summary of the VISTA-GCAV survey, which works as complementary data for our purposes. Sect. \ref{sec:photometry} describes the procedure for the source extraction, calibration of photometric quantities, and catalog construction for the cluster M0416. Sect. \ref{sec:photoz} details the computation of the photometric redshifts and the definition of the cluster membership. Sect. \ref{sec:structure} includes the computation of the red sequence of galaxies, the density field, and some galaxy properties according to local densities. Sect. \ref{sec:discussion} contains the discussion of our main results in context with the literature. Conclusions are given in Sect. \ref{sec:summary}.  

Throughout the paper, we assume a cosmology with $\Omega_m = 0.3$, $\Omega_{\Lambda}=0.7$ and $H_0=70 km s^{-1}Mpc^{-1}$. The magnitudes are given in the AB photometric system and all the distance measurements are in comoving distances. Following \citet{bal16}, the coordinates of the cluster center coincide with the NE-BCG at RA $04:16:09.14$ and DEC $-24:04:03.1$. For the virial region, we used a region of radius r$_{200} = 1.82 \pm 0.11$ Mpc, as estimated from the weak lensing by \citet{Umetsu2014}.

\section{Observations and data reduction}
\label{sec:data}

M0416 is imaged in the optical $ugri$ bands with the VLT Survey Telescope (VST) located at ESO's Paranal Observatory as part of the \textit{Galaxy Assembly as a function of Mass and Environment with VST} (P.I. A. Mercurio, hereafter VST-GAME) survey. This project is gathering observations for six massive galaxy clusters at 0.2$\lesssim$z$\lesssim$0.6 (Abell 2744, MACSJ0416-2403, Abell S1063, MACSJ0553.4-3342, PLCK G287.0+32.9, RXC J1514.9-1523) to investigate galaxy evolution down to 10$^9$ M$_{\odot}$ in stellar mass, in a wide and largely unexplored range of local densities. The uniqueness of this dataset lies in the wide field coverage ($\sim$20$\times$20 Mpc$^2$ at z=0.4), combined with long exposures (to reach the dwarf galaxy regime) used to determine the relative importance of different cluster assembly processes in driving the evolution of galaxies as a function of mass and environment. In addition, VST-GAME makes use of the near-infrared (NIR) observations taken for the same clusters in the context of the VISTA Public Survey program Galaxy Clusters At Vircam (G-CAV, P.I.: M. Nonino), which is aimed at observing 20 clusters of galaxies, covering $\sim$30 square degrees in total, in the infrared $YJKs$ bands. Thus, given the homogeneity, depth, and high-quality $u$-to-$K$ band photometry available for this cluster, we can select member galaxies via photometric redshifts, define the environment through the local galaxy density, possibly identify structures such as filaments, and investigate color-based galaxy properties at varying galaxy densities. 

To calibrate the photometric redshifts, we also took advantage of the spectroscopic data from the CLASH-VLT survey \citep{Rosati14}, which collected 4386 source spectra in a field of view (FoV) of $\sim$25$\times$25 arcmin$^2$ around the cluster center up to $\sim$2r$_{200}$, leading to the spectroscopic confirmation of $\sim$800 cluster members (see \citealt{bal16} for more details).  In this section, we describe the optical-NIR photometric data.

\subsection{VST data}
 
The VST-GAME survey, started in ESO period P99, is currently ongoing. It is carried out using 300h of the Italian INAF Guaranteed Time Observations (GTO) with OmegaCAM. M0416 was observed in P100 (0100.A-0570), P102 (0102.A-0603) and P104 (0104.A-0531) in four bands ($u$, $g$, $r$ and $i$). 

Total exposure times and full width half maximum (FWHM) of the images analyzed in this paper are reported in Tab.~\ref{table:photometry}. Image reduction, alignment, and co-adding were done using the standard procedure which includes overscan, bias, and flat corrections. After these steps, for each image, a weight map is created before an extraction step. The catalogs are then fed to Scamp \citep{Bertin2006scamp} using Gaia DR2 as a reference. The astrometric solutions are then used with Swarp \citep{Bertin2002swarp} to create the final, global stacked images. A catalog of point sources is matched to PanStarr \citep{Panstarrs1survey,Panstarrs1catalogue} to create and then apply an illumination correction map.  The VST $i$-band images are affected by fringing. We reduced images from KIDS \citep{KIDS2015PDR1}, close in time to the MJ0416 $i$-band observations, to create and subtract a fringe map.

\subsection{VISTA data}
The G-CAV survey (program 198.A-2008, P.I. M. Nonino) is a second-generation ESO public VISTA program. The survey started in P98 and was completed in P108. Table~\ref{table:photometry} reports the total exposure times and FWHMs of the NIR images analyzed in this paper.

Raw images were first corrected for non-linearity, and then darks and flats were applied. First-pass sky subtraction was then performed on an OB basis. For each image, a weight image, which also takes into account pixels flagged in darks and/or flats, is created. Thus similarly to optical data, the sources extracted are used to obtain an astrometric solution. A first stack, band per band, is then generated. \textit{SExtractor} \citep{Bertin1996} is then run to create a segmentation map, which is then dilated and converted into a 0-objects and 1-sky image. This global mask is mapped back to each chip of each reduced image, using the precise astrometric solution, to create an image per image mask, which is then used in the second run of sky estimation through the classical sliding window approach. The resulting sky-subtracted images are then used to perform the final stacks under analysis. All data are publicly accessible via the ESO Science Portal\footnote{\footnotesize \url{http://archive.eso.org/scienceportal/home?data_collection=GCAV\&publ_date=2020-12-07h}} or the Science Archive Programmatic and Tools Access\footnote{\footnotesize \url{http://archive.eso.org/programmatic/\#TAP}}. 

\begin{table}
\caption{VST-GAME and VISTA-GCAV photometry} 
\label{table:photometry}  
\centering                        
\begin{tabular}{c c c c c }       
\toprule    
\headcol Band & Exp. Time & FWHMs & Completeness \\
\headcol   & (ks) & (arcsec) & (mag) \\
\midrule  
\rowcol$u$  & 38.5 & 1.01 & 23.6\\
$g$  & 60.0 & 0.92 & 24.8\\
\rowcol$r$  & 40.5 & 0.67 & 24.4\\
$i$  & 38.0 & 0.62 & 23.6\\
\rowcol & & & \\
$Y$  & 28.8 & 0.77 & 24.4\\
\rowcol$J$  & 23.0 & 0.81 & 22.9\\
$Ks$ & 25.2 & 0.77 & 22.2\\
\hline
\end{tabular}
\end{table}

\section{Photometry and catalog construction}
\label{sec:photometry}

The VST-GAME and GCAV surveys aim to trace a large population of galaxies, from the brightest elliptical galaxies to the dwarf regime, in a wide range of cluster environments, from the crowded center to the less densely populated outskirts. The procedure adopted for the catalog construction was therefore optimized for this purpose, as described in the following.

Briefly, we tuned the source extraction for detecting both bright galaxies and also a large number of faint sources, paying particular attention to avoiding splitting large bright galaxies that show subclumps and deblending of close sources in crowded fields. Then, to have a reliable sample of galaxies, we also performed a robust separation between extended and point-like sources (hereafter the star/galaxy separation) up to the completeness limit magnitude.
Finally, we paid particular attention to spurious detections. Indeed, the internal optics of OmegaCAM at the VST often produces spikes, haloes, and ghosts (which we call in the following spurious regions) near bright stars. The photometry of sources detected inside these haloes could be affected by larger uncertainties, so we flagged these sources through the use of masks.

\subsection{Source detection}
\label{sec:sextractor}

\begin{figure}
\centering
\includegraphics[trim = 0 80 0 0 , width=\hsize]{{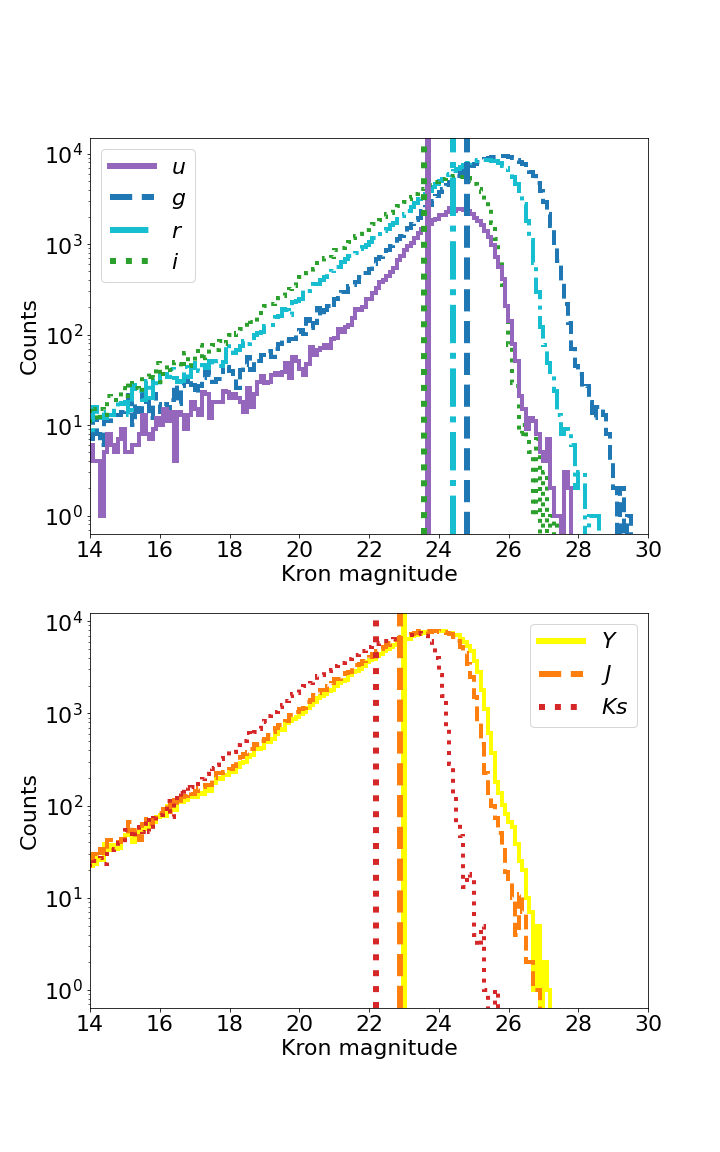}}
\caption{Magnitude distribution of sources extracted from each band in the \textit{SExtractor} single mode. The vertical lines represent the completeness limits for each band (See Table~\ref{table:photometry}).}
\label{fig:completenss}
\end{figure}

\begin{table}
\caption{Main input parameters into the \textit{SExtractor} configuration file.} 
\label{table:sexparams}  
\centering                        
\begin{tabular}{c c }       
\toprule     
Parameters & Values \\
\midrule 
\rowcol{\tt DETECT\_MINAREA} & 5 \\
{\tt DETECT\_THRESH} & 1.5 $\sigma$\\
\rowcol{\tt ANALYSIS\_THRESH} & 1.5 $\sigma$ \\
{\tt DEBLEND\_NTHRESH} & 32 \\
\rowcol{\tt DEBLEND\_MINCONT} & 0.001 \\
{\tt BACK\_SIZE} & 64 \\
\rowcol{\tt BACKFILTER\_SIZE} & 3 \\
\\
\rowcol{\tt PHOT\_APERTURES} & 1.5, 2, 3, 4, 16, 30, 45\\
 & 3*FWHM, 8*FWHM\\
\hline
\end{tabular}

\tablefoot{This setup is established for the aims of our survey, i.e. to optimize the number of individual sources, especially faint, detected in all the photometric bands. {\tt DETECT\_MINAREA} and {\tt BACK\_SIZE} are expressed in pixels. The other quantities are dimensionless and defined on \textit{SExtractor} documentation. We measured 9 {\tt PHOT\_APERTURES}, 7 are expressed in arcsecs (first row) and 2 as a function of the measured FWHM (second row).}
\end{table}

The sources are extracted using the software \textit{SExtractor} \citep{Bertin1996} together with \textit{PSFEx} \citep{Bertin2011}, which extracts precise models of the point spread function (PSF) from images processed by SExtractor. SExtractor uses the PSFEx models as input to carry out the PSF-corrected model fitting photometry for all sources in the image. The advantage of using PSF photometry is that \textit{SExtractor} can take into account the distortions of the PSF measured by \textit{PSFEx} along the field, which is particularly relevant in a large FoV. A specific validation of the PSF-corrected photometry is presented in \citet{Annunziatella2013PASP}.

\textit{SExtractor} can be used in {\it single} or in {\it dual mode}. In {\it single mode}, both the detection and the flux measurements are computed over a single image. In dual mode, the detection is done in a reference image, while the photometry is measured in a second image. As a first step, we extract the catalogs in each band ($g$,$r$,$i$,$Y$,$J$,$Ks$), except for the $u$ band, using the {\it single mode}. For the $u$ band we extract the sources directly in {\it dual mode} using the $r$ band, which has the best seeing, as the detection image. This choice was required to improve both the extraction of galaxies characterized by the presence of subclumps, which were often split into sub-components due to the nature of UV star-forming emission and the detection of faint sources in this band. Finally, we use the $r$ band as the detection image, in {\it dual mode}, to measure the photometry in the $g$, $i$, $Y$, $J$, and $Ks$ bands of sources detected in the $r$ band up to the completeness limit, but not extracted in {\it single mode} in the other bands; this is mainly because they are faint and, therefore, below the 
corresponding completeness limit (see below). 

We adopt the same setup of the \textit{SExtractor} parameters for VST images as described in \citet{Mercurio2015,Mer21} for a similar dataset, although we fix the {\tt BACK\_SIZE} to 64 pixels and {\tt BACKFILTER\_SIZE} to 3 pixels. The {\tt BACK\_SIZE} parameter determines the mesh size employed to model the background around each image pixel. The {\tt BACKFILTER\_SIZE} determines the size of the median filter to suppress possible local overestimations due to bright stars. Dealing with a large field implies that there are several kinds of objects, from point sources to diffuse galaxies, so the background level is not uniform across the field. If {\tt BACK\_SIZE} is too small, the background estimation is affected by the presence of objects and random noise. Most importantly, part of the flux of the most extended objects can be absorbed in the background map. If the mesh size is too large, it cannot reproduce the small-scale variations of the background. 

The threshold parameters ({\tt DETECT\_THRESHOLD} and {\tt ANALYSIS\_THRESHOLD}) are chosen to maximize the number of detected sources, while simultaneously keeping the number of spurious detection to a minimum. The value for the threshold is chosen according to the procedure detailed in Fig.~4 of \citet{Mercurio2015}. A summary of the key \textit{SExtractor} parameters and the aperture diameters used in this work is given in Table \ref{table:sexparams}.

We define the completeness magnitude limit of the extracted catalogs as the magnitude at which we start to lose real galaxies because they are fainter than a brightness limit. For this, we follow the method of \citet{Garilli1999} that was also adopted in \citet{Mercurio2015}. In the last column of Table~\ref{table:photometry} we report the completeness-limiting magnitude for each band, using the single-mode extractions. We compute a single-mode extraction for the $u$ band just for computing its completeness limit. 
In Fig.~\ref{fig:completenss}, we plot the magnitude distribution of the sources detected in {\it single mode} in each band. 
Considering sources within the magnitude limit proper of each band, the band catalogs include 14591 detections in the $u$ band, 78871 in the $g$ band, 84964 in the $r$ band, 64547 in the $i$ band, 103862 in the $Y$ band, 109724 in the $J$ band, and 98275 in the $Ks$ band. The larger number of sources in the NIR bands is motivated by the larger FoV of the VISTA telescope.

\subsection{Star/Galaxy separation}
\label{sec:stargalaxy}

\begin{figure*}
\centering
\includegraphics[trim= 0 50 0 50, width=\hsize]{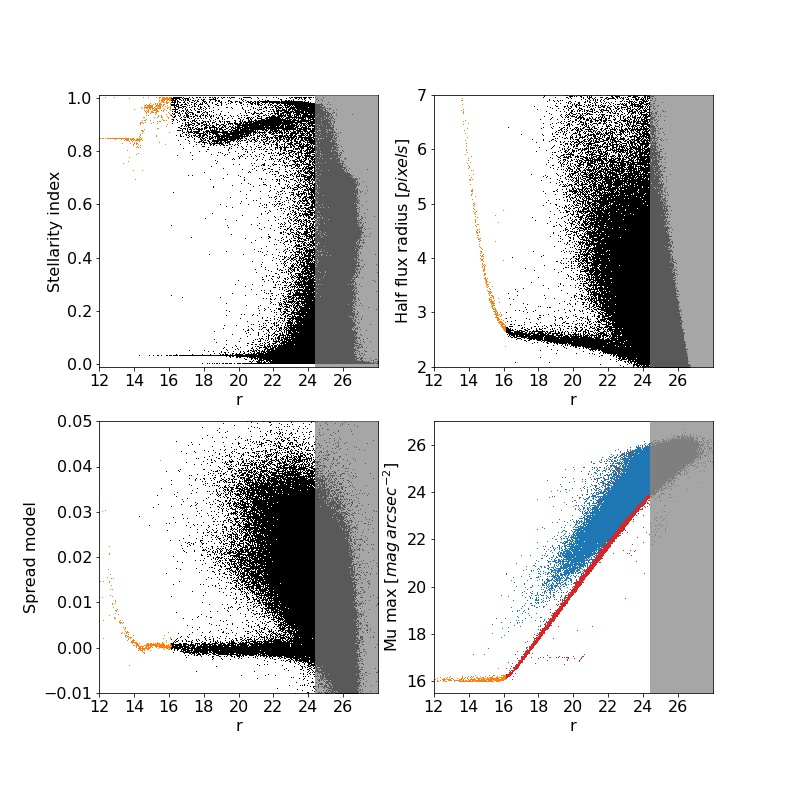}
\caption{Star/galaxy separation. Diagram of the \textit{SExtractor} {\tt CLASS\_STAR} ({\it Top-left panel}), {\tt HALF\_FLUX\_RADIUS\_50} ({\it Top-right panel}), and {\tt SPREAD\_MODEL} ({\it Bottom-left panel}), as a function of Kron magnitude for sources detected in the $r$ band (black points), with the loci of saturated stars colored in orange. The gray-shaded area corresponds to the area beyond the completeness limit of the $r$ band. In the {\it bottom-right panel} we plot the \textit{SExtractor} {\tt MU\_MAX} parameter as a function of the Kron magnitude for sources detected in the $r$ band, color-coded according to their classification: galaxies in blue, saturated stars in orange, unsaturated stars in red.}

\label{fig:star_galaxy}
\end{figure*}

To separate point-like and extended sources we adopt a procedure similar to that described in \citet{annunziatella2013}, using the following parameters: (i) the stellarity index; (ii) the half-light radius; (iii) the new \textit{SExtractor} classifier {\tt SPREAD\_MODEL}; and (iv) the peak of the surface brightness above the background ($\mu$max). The procedure is done for each band separately. 

Preliminarily, we clean the catalogs by removing spurious detections with incomplete isophotal values or corrupted memory overflow that occurred during deblending or extraction. We also remove detections with the {\tt FLUX\_RADIUS\_50} below 1 pixel, which are likely warm pixels or residuals from the cosmic ray rejection. Each step of the classification is intended to go deeper in magnitude using one \textit{SExtractor} parameter and follows the order of the four panels in Fig. \ref{fig:star_galaxy}, where black points refer to all sources, orange points are detections brighter than the saturation limit $r=$ 16.1 mag (see below), and the gray shaded region indicates those that are outside the completeness limit of the band.

In terms of the stellarity index ({\tt CLASS\_STAR}), \textit{SExtractor} identifies objects with {\tt CLASS\_STAR}=0  as galaxies and those with {\tt CLASS\_STAR}=1 as stars; so, traditionally, the star galaxy separation has been done exclusively using this parameter. However, the top left panel of Fig. \ref{fig:star_galaxy} shows the presence of multiple sequences in the {\tt CLASS\_STAR} - Kron magnitude \citep{kron80} plane for $r$-band sources, suggesting that a unique value of the {\tt CLASS\_STAR} parameter is not enough in our case. Taking advantage of the spectroscopic data we have at our disposal, we found that sources with {\tt CLASS\_STAR} $>$ 0.8 are both stars and galaxies. We then decided to consider sources under the saturation limit of $r < 16.1$ mag with a stellarity index greater than 0.5 as stars, and sources from all the magnitudes, $r\ge16.1$ mag, with a stellarity index greater than or equal to 0.99, which are surely stars. We use other \textit{SExtractor} parameters to further separate stars and galaxies.

The half light radius ({\tt FLUX\_RADIUS\_50}), is the radius in pixels containing half of the galaxy's light and can be used as a direct measure of source concentration. The top-right panel shows the {\tt FLUX\_RADIUS\_50} as a function of the Kron magnitude for $r$-band sources. Since we assume that stars are point-like sources, we can identify the stellar locus as the sequence visible in the left and lower part of the plot. This sequence remains distinguishable from the extended sources only down to approximately r$\sim$22.5 mag. Considering that the 95\% completeness limit of the $r$ band is 24.4, we have to use another parameter to separate stars from galaxies at magnitudes fainter than 22.5.

The spread classifier ({\tt SPREAD\_MODEL}), is the parameter that considers the difference between the model of the source and the model of the local point-spread function obtained with \textit{PSFex} \citep{desai2012}. By construction, it is close to zero for point sources and positive for extended sources. In the bottom left panel, the sequence of stars is visible, having \textit{{\tt SPREAD\_MODEL}}$\sim$0 as a function of the Kron magnitude. This parameter allows us to extend the classification to fainter magnitudes up to the completeness limit $r$=24.4 mag.

The peak of the surface brightness above the background ({\tt MU\_MAX}), is the value that represents the peak in the surface brightness given in magnitudes per square arcsecond. This parameter is used as a test for the star/galaxy separation computed in previous steps, and for the magnitude to separate saturated from non-saturated stars. In the bottom right panel,  {\tt MU\_MAX} is shown as a function of the Kron magnitude. The horizontal loci of saturated stars (orange) are visible, with an almost constant value of \textit{{\tt MU\_MAX}} down to $r=16.1$ mag. The sequence of stars previously identified is shown in red, and galaxies are shown in blue. The fact that stars identified in the previous steps follow a clear pattern in this panel validates our procedure.

\begin{figure}
\centering
\includegraphics[width=\hsize]{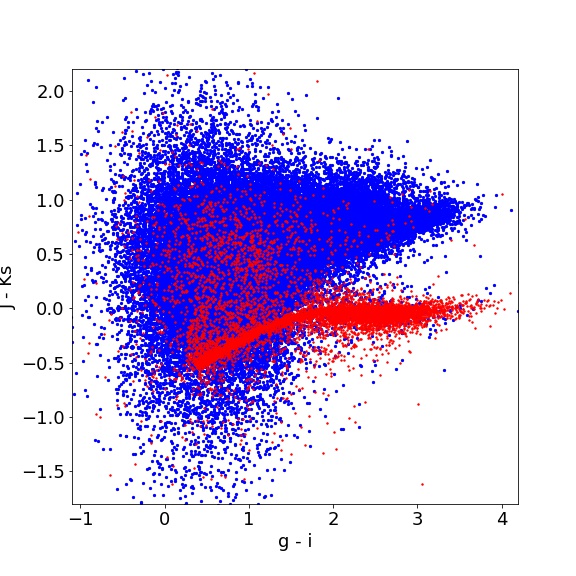}
\caption{$J \, - \, Ks$ versus $g \, - \, i$ color-color diagram, using aperture magnitudes within $3^{\arcsec}$ diameter to visualise the star/galaxy separation. Here, stars are in red and galaxies in blue. The contamination of red points on the galaxy region with $J \, - \, Ks>0.2$ is 14.8 \% and agrees with the 17.3 \% expected from the spectroscopic sample.}
\label{fig:colcol_stargal}
\end{figure}

As a further validation of our star/galaxy classification, we plot the optical versus NIR color-color diagram. According to \citet{Baldry2010} and \citet{Jarvis2013}, the locus of stars is defined by a sequence in the region with $J\,-\,Ks<0.2$, and the locus of galaxies is shown in the upper left cloud of the diagram. As shown in Fig. \ref{fig:colcol_stargal}, we found a very good agreement between the defined star/galaxy loci and the star/galaxy separation in the $r$ band up to its completeness limit. We also found that 14.8 \% of the sources classified as stars ({\tt NSFLAG\_r}>0) have $J\,-\,Ks>0.2$, which is in agreement with the spectroscopic sample. In fact, 17.3\% of the spectroscopically confirmed stars are located in the region with $J\,-\,Ks>0.2$. This also suggests that the optical-NIR color-color plot is a useful tool to separate galaxies from stars, but with residual contamination of stars of $\sim$ 15\%.

The star/galaxy separation procedure identifies 805 saturated stars, 7510 non-saturated stars, and 66323 galaxies within the completeness limit of the $r$ band. 

\subsection{Spurious regions}
\label{sec:mask}

\begin{figure}
\centering
\includegraphics[width=\hsize]{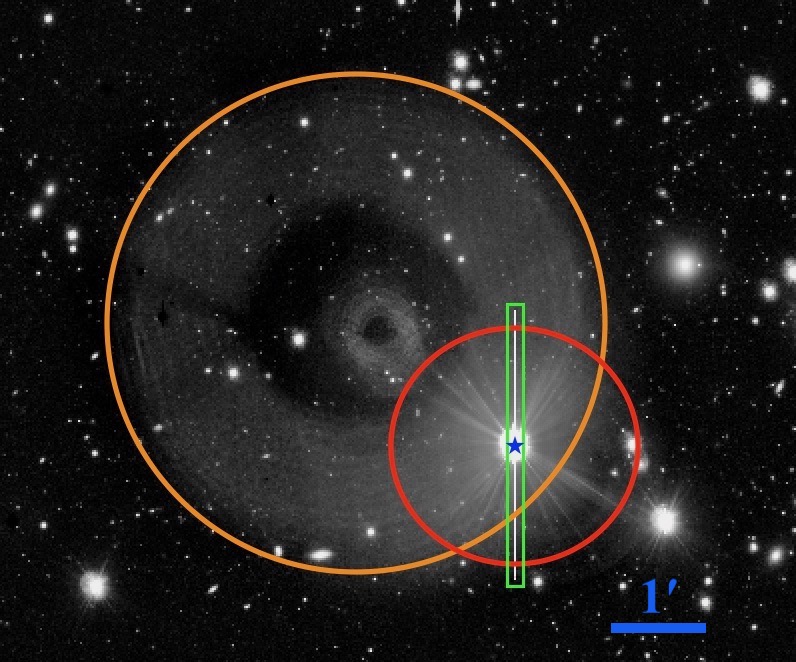}
\caption{Example of the three kinds of regions with enhanced surface brightness generated by a star in the $r$-band image; the vertical spike is shown in green, the halo centered on the star in red, and the ghost aligned in the radial direction concerning the center of the OmegaCAM field in orange. The size of the image is nearly $\mathbf{2750 \times 2150}$ pixels, which corresponds to $\mathbf{7.2' \times 9.2' }$. The blue rule indicates 1 arcmin as a reference.}
\label{fig:halo}
\end{figure}

\begin{figure*}
\centering
\includegraphics[width=\hsize]{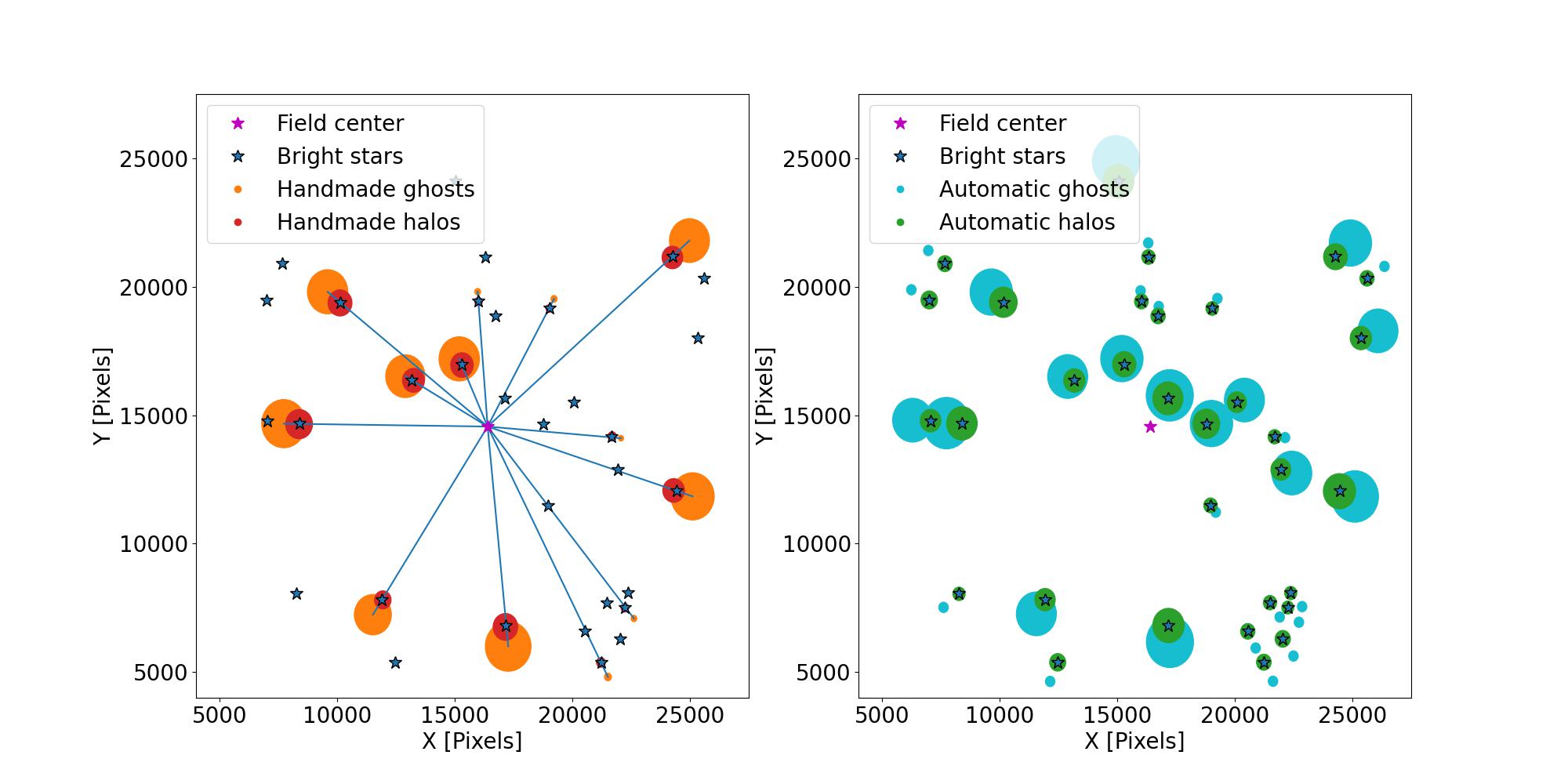}
\caption{Graphical description of the masking procedure for the OmegaCAM field. {\it Left Panel}: Star haloes and ghosts hand-drawn around bright stars in the $r$ band. The magenta star is the central pixel of the OmegaCAM field, the blue stars are all the stars with GAIA $mag \, G<13$, the red circles are the hand-drawn halos centered around the corresponding star, and the orange circles are the hand-drawn ghosts placed in a radial direction concerning the field center. Hand-drawn haloes and ghosts are generated for a limited sample of the bright stars. Blue lines connect the center of the ghosts with the center of the field. Using the sample of hand-drawn haloes and ghosts (position, diameter, and magnitude of the star), we built a mask for all the bright stars.
{\it Right Panel}: Mask for all the stars brighter than GAIA $mag \, G<13$. The procedure to generate the mask is calibrated from the regions visually identified in the left panel and applied to all the bright stars in the field. On both panels, north is up and east is left.}
\label{fig:mask}
\end{figure*}

As already mentioned, the presence of bright stars in the observed images can produce complex regions with enhanced surface brightness due to multiple reflections of the light in the internal optics of OmegaCAM. A visual inspection of the sources lying in and around these regions suggests that these sources might have their flux measurements affected by these irregular patterns, it is therefore very important to properly identify them.

There are three main kinds of impurities due to bright stars inside the field: spikes, halos, and ghosts, an example of which is shown in Fig.~\ref{fig:halo}. 
Spikes are narrow bright lines coming out from the star vertically (see the green rectangle in Fig.~\ref{fig:halo}). Haloes are circular regions with higher luminosity centered on the star (see the red circle in Fig.~\ref{fig:halo}). Ghosts are composed of a central region with low surface brightness surrounded by an outer corona with magnified surface brightness characterized by an irregular pattern. These features are not centered on the star that generated them, but outside in a radial direction concerning the center of the OmegaCAM field (see the orange circle in Fig.~\ref{fig:halo}). 

We developed a tool to mask haloes and ghosts using the position and magnitude of the stars that generate them. Spikes are almost always contained inside haloes and involve no sources, so there is no need to mask them.  
We start with the visual inspection of the $r$-band image and verify that indeed all the haloes and ghosts are generated by saturated stars. We compare with the Gaia catalog \citep{gaia_dr2} and find that stars with a $G$ magnitude brighter than 12 generate the strongest haloes and ghosts. Stars with $12<G<13$ still generate small spurious regions with a low noise level and affect only a few detected objects. Thus, we draw circular regions close to the stars brighter than $G \sim 13$ to adopt a conservative approach. By studying the size and distribution of some of these visually selected regions, shown in the left panel of Fig. \ref{fig:mask}, we implement a general procedure to automatically generate circular masks covering haloes and ghosts, as detailed in Appendix \ref{appendix:a}. The final mask for the $r$ band is shown in the right panel of Fig. \ref{fig:mask}. The mask covers 5.29\%, 12.29\%, 12.93\%, and 9.85\% of the 1 deg$^2$ VST FoV in the $u$, $g$, $r$ and $i$ bands, respectively. The mask covers 2.20\%, 2.84\%, and 1.73\% of the 1.5 deg$^2$ VISTA FoV in the $Y$, $J$, and $Ks$ bands, respectively. Regardless of the independent masking of every single band, in the matched catalog we consider the $r$ band to be the reference masking image for all the objects in the field.

Additionally, due to the \textit{SExtractor} setup adopted for this work, we find a certain number of sources following strange patterns that coincide with the geometry of the spurious regions inside the mask. Those sources are considered fake detections and there are nearly 0.4\% of them in the single-band catalogs within the completeness limit and located inside the mask. Fake detections were removed manually from single-band catalogs. In the multi-band catalog (see Sec. \ref{sec:multiband}), the issue of spurious detections is avoided since the probability of simultaneously detecting a fake object over seven bands is low because the geometry of spurious regions is different for each band. Fake objects follow different patterns, which are canceled by the matching band.

After the masking procedure on the multiband sample (see Sec. \ref{sec:multiband}), we find that 7169 sources ($9.6\%$) are inside the mask ({\tt MASK\_FLAG}=1), while 66945 sources ($90.4\%$) are outside the mask ({\tt MASK\_FLAG}=0).

\subsection{Multiband catalogue}
\label{sec:multiband}

\begin{sidewaystable*}
\caption{Extract of the first lines of the catalog. } 
\label{table:firstrows}  
\centering

\begin{tabular}{@{}c c c c c c c c@{}}       
\toprule
\headcol GAME\_ID & RA & DEC & A & B & THETA & R\_50 & R\_KRON \\
\headcol  & [deg] & [deg] & [pixel] & [pixel] & [deg] & [pixel] & [pixel] \\
\midrule
\rowcol GAME0414225-2436203 & 63.593922 & -24.605635 & 28.11 & 23.52 & -52.19 & 22.95 & 3.50  \\
GAME0414248-2436296 & 63.603519 & -24.608214 & 10.91 & 7.38 & -84.86 & 7.80 & 3.50 \\
\rowcol GAME0416207-2436258 & 64.086120 & -24.607161 & 16.08 & 8.37 & 82.97 & 12.16 & 3.50 \\
GAME0415237-2436535 & 63.848645 & -24.614866 & 19.49 & 9.47 & -31.09 & 10.70 & 3.50 \\
\rowcol GAME0416001-2436511 & 64.000477 & -24.614206 & 7.08 & 6.27 & -66.84 & 3.46 & 3.50 \\
GAME0414148-2437017 & 63.561682 & -24.617152 & 3.07 & 1.54 & 19.05 & 3.79 & 4.39 \\
\rowcol GAME0414408-2437028 & 63.670192 & -24.617455 & 2.53 & 1.31 & 3.95 & 3.44 & 4.49 \\
GAME0414229-2437001 & 63.595250 & -24.616692 & 4.04 & 2.49 & 89.69 & 4.42 & 3.53 \\
\rowcol GAME0415287-2437023 & 63.869606 & -24.617305 & 2.58 & 2.14 & -31.40 & 2.78 & 3.51 \\
\hline
\end{tabular}

\begin{tabular}{@{}c c c c c c c c c c@{}}       

\toprule
\headcol GAME\_ID & AP\_15\_$u$ & APERR\_15\_$u$ & AP\_30\_$u$ & APERR\_30\_$u$ & AP\_40\_$u$ & APERR\_40\_$u$ & Mag\_Kron\_$u$ & MagERR\_Kron\_$u$ \\
\headcol   & [mag] & [mag] & [mag] & [mag] & [mag] & [mag] & [mag] & [mag] \\
\midrule

\rowcol GAME0414225-2436203 & 21.63 & 0.06 & 20.26 & 0.06 & 19.72 & 0.06 & 17.86 & 0.11 \\
GAME0414248-2436296 & 15.20 & 0.06 & 14.60 & 0.07 & 14.48 & 0.07 & 14.33 & 0.11\\
\rowcol GAME0416207-2436258 & 22.39 & 0.07 & 21.15 & 0.07 & 20.75 & 0.07 & 19.60 & 0.09\\
GAME0415237-2436535 & 23.35 & 0.11 & 22.35 & 0.10 & 22.08 & 0.10 & 21.21 & 0.19 \\
\rowcol GAME0416001-2436511 & 17.51 & 0.06 & 16.94 & 0.06 & 16.83 & 0.06 & 16.71 & 0.07\\
GAME0414148-2437017 & 25.02 & 0.51 & 23.94 & 0.43 & 23.61 & 0.42 & 23.54 & 0.39 \\
\rowcol GAME0414408-2437028 & 24.93 & 0.48 & 24.41 & 0.63 & 24.34 & 0.76 & 24.43 & 0.69 \\
GAME0414229-2437001 & 25.04 & 0.47 & 23.79 & 0.32 & 23.79 & 0.41 & 23.75 & 0.44 \\
\rowcol GAME0415287-2437023 & 25.55 & 0.75 & 24.46 & 0.63 & 23.93 & 0.52 & 24.05 & 0.49 \\
\hline
\end{tabular}

\begin{tabular}{@{}c c c c c c c c c c c@{}}       

\toprule
\headcol GAME\_ID & Mag\_Model\_$u$ & MagERR\_Model\_$u$ & Mag\_PSF\_$u$ & MagERR\_PSF\_$u$ & A\_$u$ & [...] & SI & NSFLAG & MASK\_FLAG & photo\_$z$ \\
\headcol   & [mag] & [mag] & [mag] & [mag] & [mag] & - & - & - & - & -\\
\midrule

\rowcol GAME0414225-2436203 & 17.86 & 0.12 & 21.46 & 0.06 & 0.188 & (...) & 0.03 & 0.0 & 0.0 & 0.17 \\
GAME0414248-2436296 & 14.31 & 0.16 & 14.89 & 0.07 & 0.192 & (...) & 0.75 & 9.0 & 0.0 & -99.00\\
\rowcol GAME0416207-2436258 & 19.60 & 0.07 & 22.29 & 0.07 & 0.193 & (...) & 0.03 & 0.0 & 0.0 & 0.30\\
GAME0415237-2436535 & 21.21 & 0.10 & 23.23 & 0.10 & 0.193 & (...) & 0.03 & 0.0 & 0.0 & 0.33\\
\rowcol GAME0416001-2436511 & 16.64 & 0.11 & 17.23 & 0.06 & 0.186 & (...) & 0.97 & 9.0 & 0.0 & -99.00\\
GAME0414148-2437017 & 22.99 & 0.55 & 24.60 & 0.37 & 0.193 & (...) & 0.42 & 0.0 & 0.0 & 0.45\\
\rowcol GAME0414408-2437028 & 24.36 & 1.68 & 24.79 & 0.45 & 0.186 & (...) & 0.98 & 0.0 & 0.0 & 0.38\\
GAME0414229-2437001 & 23.75 & 0.65 & 24.67 & 0.36 & 0.186 & (...) & 0.03 & 0.0 & 0.0 & 0.18\\
\rowcol GAME0415287-2437023 & 21.60 & 4.23 & 25.05 & 0.49 & 0.185 & (...) & 0.96 & 0.0 & 0.0 & 0.32\\
\hline
\end{tabular}
\tablefoot{All the quantities with the subindex $\_u$ in this table are present for all the bands in the place of the [...] column. The other quantities are reported from the $r$ band, and the photo-$z$ includes information in all bands. }
\end{sidewaystable*}

With this paper, we publish the multiband (optical-NIR) catalog, obtained by matching all the sources in the $r$-band catalog within the completeness limit of $r<$ 24.4, using {\tt STILTS} \citep{taylor2006}. Detections in the $u$, $g$, $i$, $Y$, $J$, and $Ks$ bands within a 1$^{\arcsec}$ distance from $r$-band sources are identified as a match. Magnitudes in the $ugiYJKs$ bands can be fainter than their corresponding completeness limit since we do not cut these catalogs before the match. 
In the catalog, we report a unique primary key ({\tt GAME\_ID}), which univocally identifies sources. It is composed of a string containing 19 characters, where four of them are ``GAME'' and fourteen for the digits of the hhmmsss-ddmmsss barycentre coordinate, which we report in degrees ({\tt RA}, {\tt DEC}). We also report the parameters of the ellipse that describes the shape of the objects: the semi-major and semi-minor axes ({\tt A} and {\tt B}), and position angle ({\tt THETA}), together with the half flux radius ({\tt R\_50}) and the Kron radius ({\tt R\_Kron}) measured in the $r$ band. Among all the measured magnitudes, we report three aperture magnitudes inside a 1.5, 3, and 4 arcsec diameter ({\tt AP\_15, AP\_30, AP\_40}), the Kron magnitude ({\tt Mag\_Kron}) and the model magnitude obtained from the sum of the spheroid and disk components of the fitting ({\tt Mag\_Model}), as well as the PSF magnitude ({\tt Mag\_PSF}). Magnitudes are corrected for the galactic extinction according to \citet{Schlafly2011}, and we also report the adopted corrections ({\tt A\_u, A\_g, A\_r, A\_i, A\_Y, A\_J, A\_Ks}) in the catalog. Additionally, to obtain a more realistic photometric error, we multiply the nominal error given by \textit{SExtractor} by the factors given in Table 6 of  \citet{Mercurio2015}: $R=a_1 \times \exp[a_2\times({\rm mag}-20)]+a_3$, where mag is the magnitude of the source and the $a_i$ coefficients are calibrated for each band. For the VISTA magnitudes, we use the same coefficients as for the $i$ band. 

Finally, we provide the stellarity index obtained for the $r$ band from \textit{SExtractor} ({\tt SI}) and two additional flags: the star/galaxy flag ({\tt NSFLAG\_r}) and the mask flag ({\tt MASK\_FLAG\_r}). {\tt NSFLAG\_r} flag is set to 0 for extended objects and greater than zero for point-like sources identified in the $r$-band, according to the star/galaxy separation described in Sect.~\ref{sec:stargalaxy}.  {\tt MASK\_FLAG\_r} is set to one for objects inside the $r$ band mask and zero otherwise, according to the identification of star haloes described in Sect.~\ref{sec:mask}. We also report the measured photometric redshift ({\tt photo\_$z$}), whose determination is presented in the next section.

The multiband catalog contains 74114 sources. 
Table \ref{table:firstrows} contains a general view of the first rows of the multiband catalog and Table \ref{table:cataloguestatistics} provides some statistics on the number of objects.
The single-band catalogs, including the star/galaxy separation and the masking calibrated for each band, are available upon request.

\begin{table}
\caption{Number of objects in the multiband catalog.
} 
\label{table:cataloguestatistics}  
\centering                        
\begin{tabular}{c c}       
\toprule       
\textit{Total number of objects} & 74 114 \\
\midrule
\rowcol Galaxies ({\tt NSFLAG\_r}=0) & 66 323\\
Stars ($0<${\tt NSFLAG\_r}$<9$) & 7 510\\
\rowcol Saturated stars ({\tt NSFLAG\_r}$\ge9$) & 805 \\
\\
Non masked objects ({\tt MASK\_FLAG\_r}=0) & 66 945\\
\rowcol Masked objects ({\tt MASK\_FLAG\_r}=1) & 7 169\\
\\
Objects without \textit{zspec} & 69 991 \\
\rowcol Objects with \textit{zspec} (cluster core) & 4 123 \\
\hline
\end{tabular}
\tablefoot{The table includes a classification according to star/galaxy separation, masking, and the availability of spectroscopic redshift.}
\end{table}

\section{Photometric redshift and cluster membership}
\label{sec:photoz}

The quality and depth of the multiband catalog, introduced in Section \ref{sec:multiband}, allow us to assign accurate photometric redshifts to the sources detected in the VST field centered on the M0416 cluster. In this section, we describe the iterative procedure adopted to tune the photometric redshift determination and a description of the criteria adopted to define the cluster membership.

\subsection{Photometric redshifts}

\begin{figure}
\centering
\includegraphics[width=\hsize]{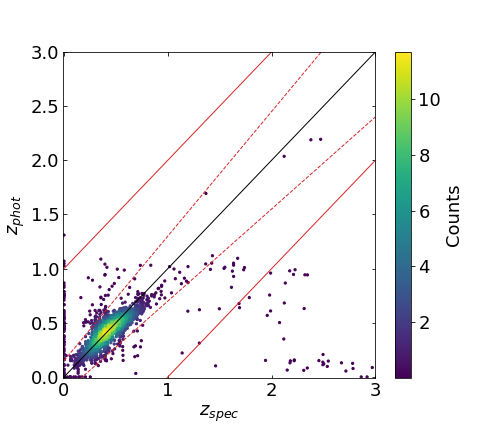}
\caption{Results of the SED fitting procedure, before optimizing it into our redshift range. The figure shows the photometric redshift as a function of spectroscopic redshift. The internal red dotted lines indicate the limit for which $|\Delta z|=\mathbf{|(z_{phot} - z_{spec})/(1 + z_{spec})|}>0.15$, which defines the outliers. Red continuos lines indicate the limit for catastrophic outliers $|z_{phot}-z_{spec}|>1$. The outlier fraction is 6.06\% and the $\sigma_{NMAD}$ is 0.0418.}
\label{fig:first_photoz}
\end{figure}

Photometric redshifts are determined through a spectral energy distribution (SED) fitting approach, using the code \textit{LePhare} \citep{Arnouts1999,Ilbert2006}.
We perform the first run on the multi-band catalog, using all the available bands ($u$, $g$, $r$, $i$, $Y$, $J$, $Ks$) and testing the performance of the fitting using different magnitudes. In particular, we test three aperture magnitudes inside $1.5 \arcsec$, $4.0 \arcsec$, and $8.0 \arcsec$ diameter, the Kron magnitude, and the model magnitude, which takes into account and correct for the PSFs obtained for each image. We find that the most accurate results are obtained using the smallest aperture magnitude in our catalog, $1.5 \arcsec$. This fact confirms the results of \citet{Jouvel2014}, who found that the optimal \textit{LePhare}'s performance is obtained with small aperture magnitudes. Since we are using PSF fitting photometry, the aperture magnitudes already take into account the differences in seeing among the bands, so the fraction of the source's flux is the same for a given aperture in all the bands.

To calibrate our photometric measurements and quantify their accuracy, we benefit from the rich spectroscopic sample available for M0416 \citep{bal16,Caminha17,van19}, covering the area at r<2 r$_{200}$. 
We match the spectroscopic sample with our multi-band catalog and we identify 4123 objects with both VST-GAME photometry and CLASH-VLT spectroscopy.
The sources with spectroscopic redshift allow us to compute zero point corrections for each photometric band through the {\tt AUTO\_ADAPT} parameter on \textit{LePhare}. 

We, thus, obtain the photo-$z$ measurements presented in Fig. \ref{fig:first_photoz}, where we report the distribution of photometric versus spectroscopic redshifts. To quantify the goodness of the photometric redshifts, we use the quantity $\Delta z = (z_{phot} - z_{spec})/(1 + z_{spec})$ following \citet{Euclid2020}. Objects with $|\Delta z| > 0.15$ are regarded as outliers, and their fraction, $\eta$, indicates the accuracy of the fitting. After excluding the outliers, we compute the following photo-$z$ statistical estimators:
the mean of $|\Delta z|$, \textit{bias}; 
the standard deviation of $|\Delta z|$, $\sigma$; 
and the standard deviation of the normalized median absolute deviation, $\sigma_{NMAD}$, given by $1.4826 \times \textit{median}(|\Delta z|)$ \citep{Razim2021,Hong2022}.
The quality of results shown in Fig. \ref{fig:first_photoz} can be described with $bias = 0.0358$, $\sigma = 0.0296$, $\sigma_{NMAD}=0.0418$ and $\eta=6.06 \%$.  

\begin{figure}
\centering
\includegraphics[width=\hsize]{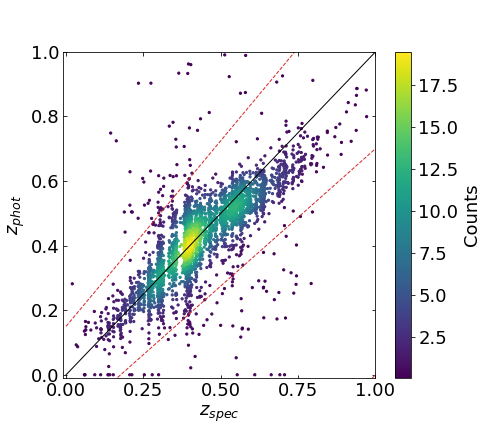}
\caption{Photometric redshift as a function of the spectroscopic redshift optimized at the cluster interval. This SED fitting is calibrated with a sample of galaxies which excludes stars ({\tt NSFLAG\_r}>0 or $z_{spec}=0$) and galaxies with $z_{spec}>1$. The fraction of outliers is 3.75\% and the $\sigma_{NMAD}$ is $0.0411$.}
\label{fig:photoz}
\end{figure}

We can note that the focus of this work is the analysis of cluster member galaxies, so to improve the performance of the fitting procedure and focus it on the cluster redshift, we select only the galaxies ({\tt NSFLAG\_r}=0) and for those with spectroscopy, we impose $0<z_{spec}<1$. Thus avoiding also the problem of the spectroscopic under-sampling of the sources at $z>1$.

We obtain a final photometric sample of 66211 galaxies and a spectroscopic sample of 3840 galaxies. The new fitting run is shown in Fig. \ref{fig:photoz}, where we compare the photo-$z$ measurements with spectroscopic redshifts. 

\begin{figure}
\centering

\includegraphics[width=\hsize]{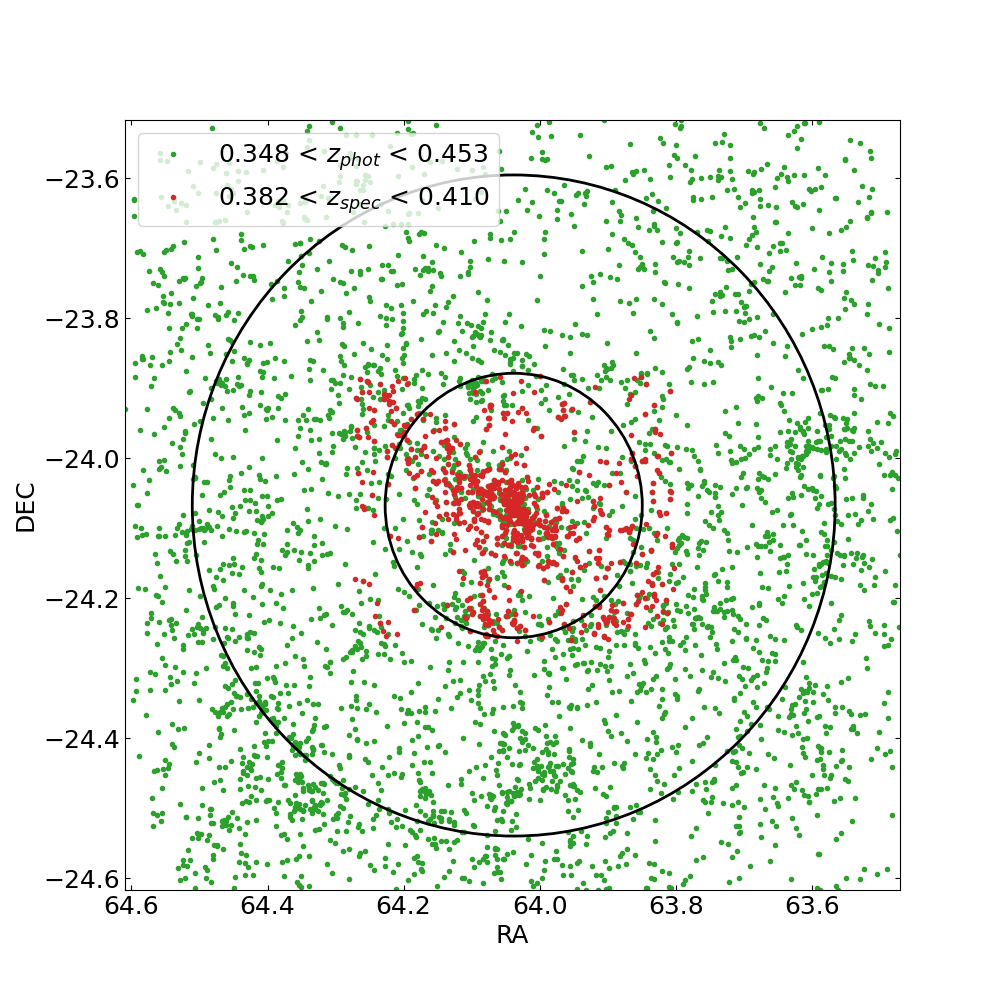}
\caption{Distribution across the VST field of spectroscopic (red) and photometric (green) cluster members. The inner circle indicates 2r$_{200}=3.64 \, Mpc$ and the outer circle indicates 5r$_{200}=9.1 \, Mpc$.}

\label{fig:hist_allphotoz}
\end{figure}

The statistics of the run are $bias = 0.0361$, $\sigma = 0.0298$, $\sigma_{NMAD} = 0.0411$, $\eta = 3.75\%$, confirming that we can use these measurements as the best photo-$z$s for our sample. Fig. \ref{fig:hist_allphotoz} shows the overall redshift distribution for our sample, highlighting the photo-$z$ and spec-$z$ data sets.

We would like to stress that among the preliminary test, we make a photo-$z$ test using the ({\tt Mag\_Model}), which includes a correction for the PSFs in each band, but the results are less accurate, leading to an outlier fraction of $\eta>15\%$. We also check the impact of a bad signal-to-noise (S/N) ratio on the NIR bands ($Y$, $J$, $Ks$). When removing NIR sources with $S/N<5$, $LePhare$ returns an improvement on the photo-$z$ accuracy, reaching a $bias=0.0354$, $\sigma=0.0293$, $\sigma_{NMAD}=0.0406$ and $\eta=3.16\%$, so in the following analysis we use only galaxies with $S/N\ge5$ in the NIR bands (N$_{\mathrm{gal}}$=49779) to enhance the robustness of our results. However, we release the photometric redshifts for all the galaxies, without considering any cut in signal-to-noise. 

\subsection{Cluster membership}
\label{sec:Cluster_membership}
\begin{figure}
\centering
\includegraphics[width=\hsize]{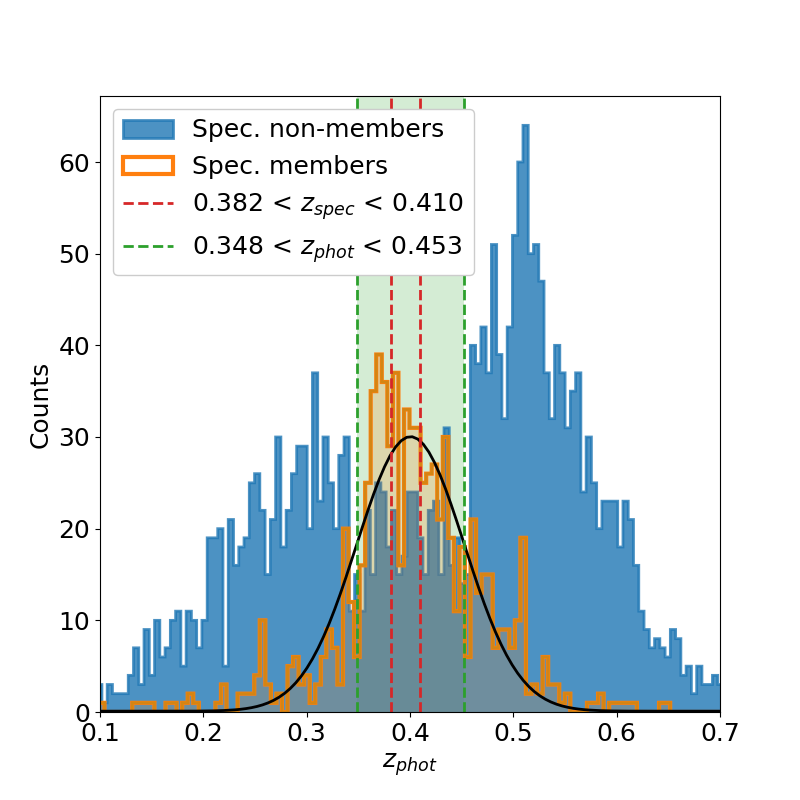}
\caption{Histogram showing the distribution of photometric redshift for objects with spectroscopic information. In orange are shown the spectroscopic cluster members, described by a Gaussian curve (black line). The photometric cluster interval is shown in green and corresponds to the one-sigma dispersion of the Gaussian curve $\sigma=0.052$. In blue are shown the objects with a spectroscopic redshift outside the cluster interval.}
\label{fig:hist_photoz}
\end{figure}

We compute the cluster memberships using spectroscopic redshifts, when available and complement the sample with photometric redshifts over the whole field. We use the spectroscopic information to optimize the interval of photometric redshifts to define the membership.

Figure \ref{fig:hist_photoz} shows the histogram of photometric redshifts for the objects in our catalog with spectroscopic information and with a $S/N>5$ in the NIR bands. In orange are shown the photo-$z$ of the 801 spectroscopic cluster members within the interval $0.382<z_{spec}<0.410$ \citep[red dotted lines in the figure, according to Fig. 5 of ][]{bal16}, which corresponds to a rest-frame velocity of $\pm 3000$ km~s$^{-1}$. In blue are shown the photo-$z$ of 2494 objects with a spectroscopic redshift outside the cluster interval. To define the photometric cluster membership, we fit a Gaussian curve over the photo-$z$ histogram of spectroscopic cluster members and select the one sigma interval, which leads us to $0.348<z_{phot}<0.453$ (green shaded area in the figure). The number of photometric cluster members is 4067 galaxies across the whole VST field. A larger interval, in terms of Gaussian sigmas, would return a larger number of cluster members (increasing the completeness, see below), nevertheless, it would include a large number of non-cluster members into the cluster interval (decreasing significantly the purity, see below). The one-sigma interval is an optimal choice to balance between completeness and purity.

We quantified the completeness and purity of the selected photometric members. The completeness is defined as the ratio between the number of galaxies identified as members (both spectroscopic and photometric) and the number of spectroscopic members $C_M = N_{pm \cap zm}/N_{zm}$. The purity is defined as the ratio between the number of photometric members that are also confirmed by spectroscopy and the number of photometric members that have $z_{spec}$, $P = N_{pm \cap zm}/N_{pm \cap z}$. The completeness of the sample of cluster members at the interval is 63.3 \% and the purity is 55.8 \%.

\section{Environment classification}
\label{sec:structure}

In the previous sections, we present the photometric catalog for M0416. We are now in the position of characterizing the properties of the galaxies in the cluster. In Sect. \ref{sec:rs} we identify the red sequence galaxies, and in Sect. \ref{sec:overdensity} we present the characterization of the density field. These two ingredients allow us to study the variation of galaxy colors in relation to the environment (Setcs. \ref{sec:localenv} and \ref{sec:outskirtstruct}).

\subsection{Red sequence galaxies}
\label{sec:rs}

\begin{figure}
\centering
\includegraphics[width=\hsize]{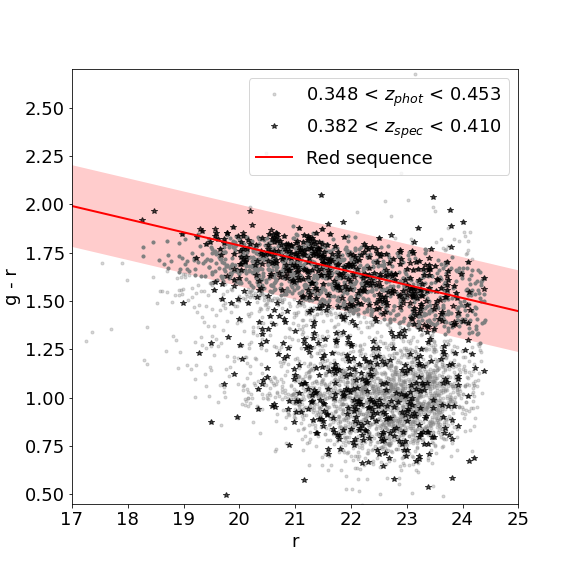}
\caption{Color magnitude diagram for MACS0416 cluster members using VST photometry. Color is determined using $g$ and $r$ band aperture magnitudes of 3.0$\arcsec$. The x-axis shows the Kron magnitude for the $r$ band. The sequence of red galaxies is determined with spectroscopic cluster members (black stars in the figure), first splitting the sample in two and then fitting a linear relation in the upper part of the plane. The red line determines the best fit for red galaxies and the red shadow area represents the $2.6 \sigma$ region around the fit. Grey points are galaxies with photometric redshift within the cluster interval.}
\label{fig:colmag}
\end{figure}

Since the population of red galaxies dominates the densest regions of large-scale structures, they are efficient tracers of the cosmic web. In this section, we present the analysis of the color-magnitude diagram used to select red galaxies across the field and to identify overdensities. 

The color-magnitude plane of all members is shown in Fig. \ref{fig:colmag}. We obtain the red sequence relation using spectroscopically confirmed cluster members ($0.382<z_{spec}<0.410$) in the  $g-r$ color within a $3 \arcsec$-diameter aperture, and the $r$-band total magnitude. Exploiting the software presented in sect. 3.2 of \citet{Cappellari2013}, we first fit a linear relation over the whole sample. Then, a subsample of galaxies with $g-r\, >-0.0985\times(r-21.99)+1.386$ are chosen to determine the red sequence relation: $g-r\,=-0.0662\times(r-21.86)+1.6631$, with a standard deviation of $\sigma=0.0813$. Finally, we define all photo-$z$-based members contained within $2.6$ times the standard deviation of this sequence as red sequence galaxies. We expect that $\sim$99\% of red galaxies included are within this confidence level. 

Considering the sample of red galaxies in the cluster redshift interval of $0.348 < z_{phot} < 0.453$, we obtain a completeness level of 100.0\% and a purity of 71.1\% concerning the sample of spectroscopic cluster members within the red sequence. This improvement confirms that red galaxies have a more accurate photometric redshift than the whole sample and that can be used as an efficient tracer of dense structures at the redshift that we are studying.

In the following, we use the red-sequence galaxies to compare galaxy populations across environmental density (Sec. \ref{sec:localenv}) and to identify the most significant overdense regions in the cluster outskirts (Sec.\ref{sec:outskirtstruct}).

\subsection{Density field}
\label{sec:overdensity}

\begin{figure*}
\centering
\includegraphics[width=14cm]{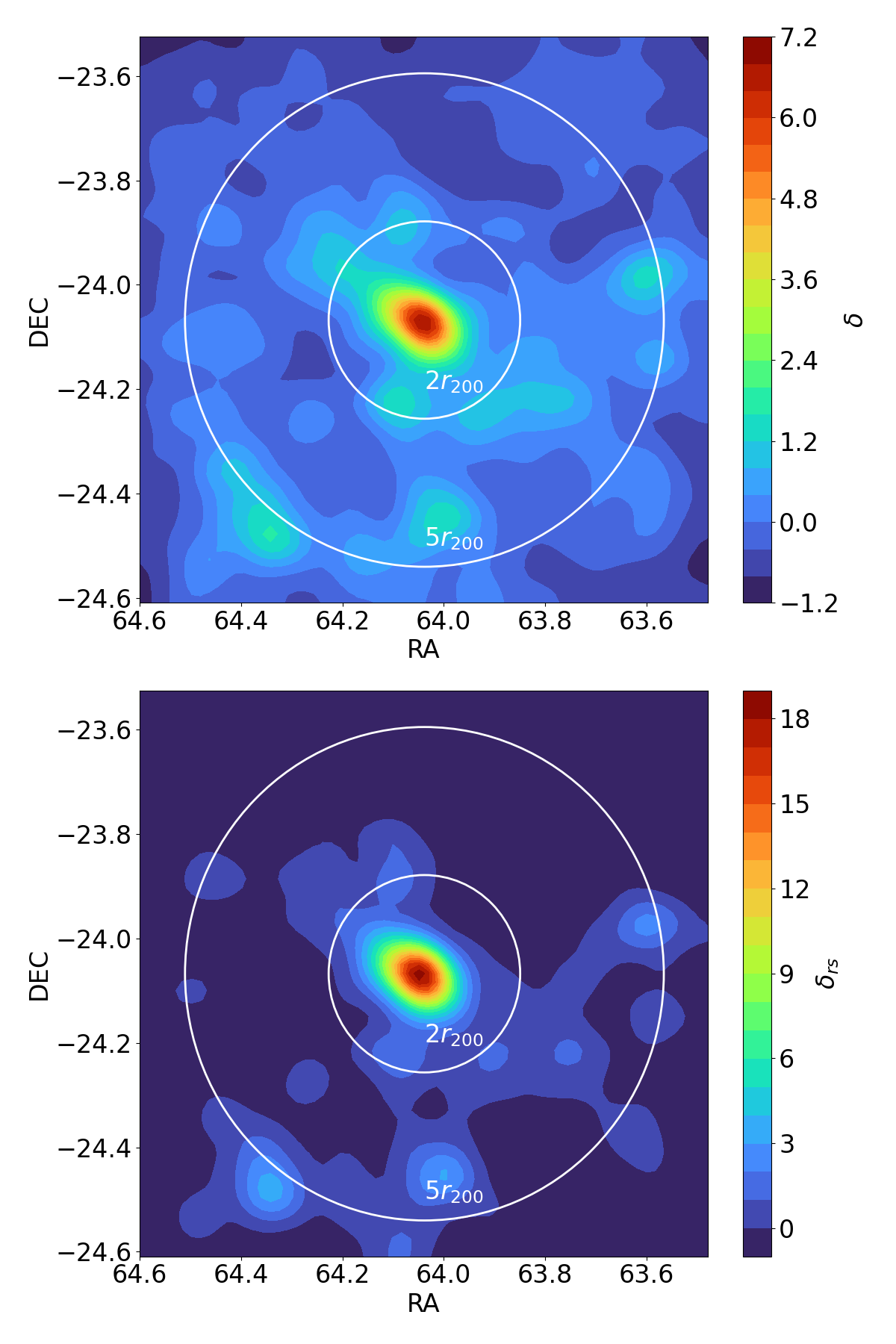}
\caption{MACS0416 density field computed using the VST-GAME photometry in the $0.348 < z_{phot} < 0.453$ redshift interval. \textit{Top:} density field computed with all galaxies in the redshift interval. The inner circle indicates 2r$_{200}=3.64 \, Mpc$ and the outer circle indicates 5r$_{200}=9.1 \, Mpc$. Both circles are centered around the NE-BCG, which coincides with the center of the cluster \citep{bal16}.
\textit{Bottom:} density field computed with red galaxies in the redshift interval.}
\label{fig:radec_zphot}
\end{figure*}

A way to determine the different environments affecting galaxies and their properties is to compute the galaxy density field \citep[e.g.,][]{Gomez2003,Kovac2010,ann14,Granett2015,ann16,Malavasi2017,Gargiulo2019,Ata2021}. We derived the projected density field using the galaxies lying in the photometric redshift interval of $0.348<z_{phot}<0.453$. Initially, we set a rectangular grid using square cells of 1 arcmin a side, corresponding to a comoving distance of $\sim$300 kpc at the cluster redshift. On this grid, we use the nearest-grid-point (NGP) scheme to count the number of galaxies in a given cell, $n_i$. 

Then, we compute the fluctuations over the mean value for each cell, $\delta_i^* = n_i/\bar{n} -1$, where $\bar{n}=0.92$ is the mean number of galaxies per cell across the whole field at the cluster redshift. Additionally, $\delta_i^*$ is convolved with a Gaussian kernel to connect the local density with adjacent cells, generating a stronger environment tracer, which takes into account not only the cell density but also the density of nearby cells. The standard deviation of the Gaussian kernel corresponds to two cells of the grid. This smoothed density field on a particular cell (for simplicity $\delta$) will be our environment tracer, and its value is assigned to each galaxy inside the cell. Additionally, we compute the density field using only red galaxies $\delta_{rs}$ following the same procedure ($\bar{n}_{rs}=0.31$).

In general, the numerical values of the density field are dependent on the method used to compute it, and, in our case, on the cell size and width of the Gaussian kernel. Cosmological studies including those of \citet{Monaco1999}, \citet{Kitaura2009}, \citet{Schirmer11}, \citet{Granett2015}, and \citet{Ata2021} used a grid size between 1 and 5 Mpc~h$^{-1}$ and Gaussian kernels up to 10 Mpc to better identify structures over large regions of the sky, while studies including those of \citet{ann14} and \citet{Malavasi2017} that were focused on galaxy evolution, on scales from $300$ kpc up to 2 Mpc to reconstruct the local environment. In our case, the cell size and width of the Gaussian kernel are chosen to find an optimal trade-off between large-scale structures and small-scale number-count fluctuations.    

In Fig. \ref{fig:radec_zphot}, we show the galaxy density field for M0416 above the $r$-band completeness magnitude limit of $r<24.4$. The top panel is the density field, $\delta$, computed with all the galaxies in the redshift interval, while the bottom panel is the density field computed using only red galaxies $\delta_{rs}$. The former is used in Sect. \ref{sec:localenv} to study populations of galaxies across several levels of environmental density, while the latter is used in Sect. \ref{sec:outskirtstruct} to detect overdense regions in the cluster outskirts at r>2r$_{200}$. As expected, in both panels it appears evident that the highest density values correspond to the central region of the cluster. 

\begin{figure*}
\centering
\includegraphics[width=19cm]{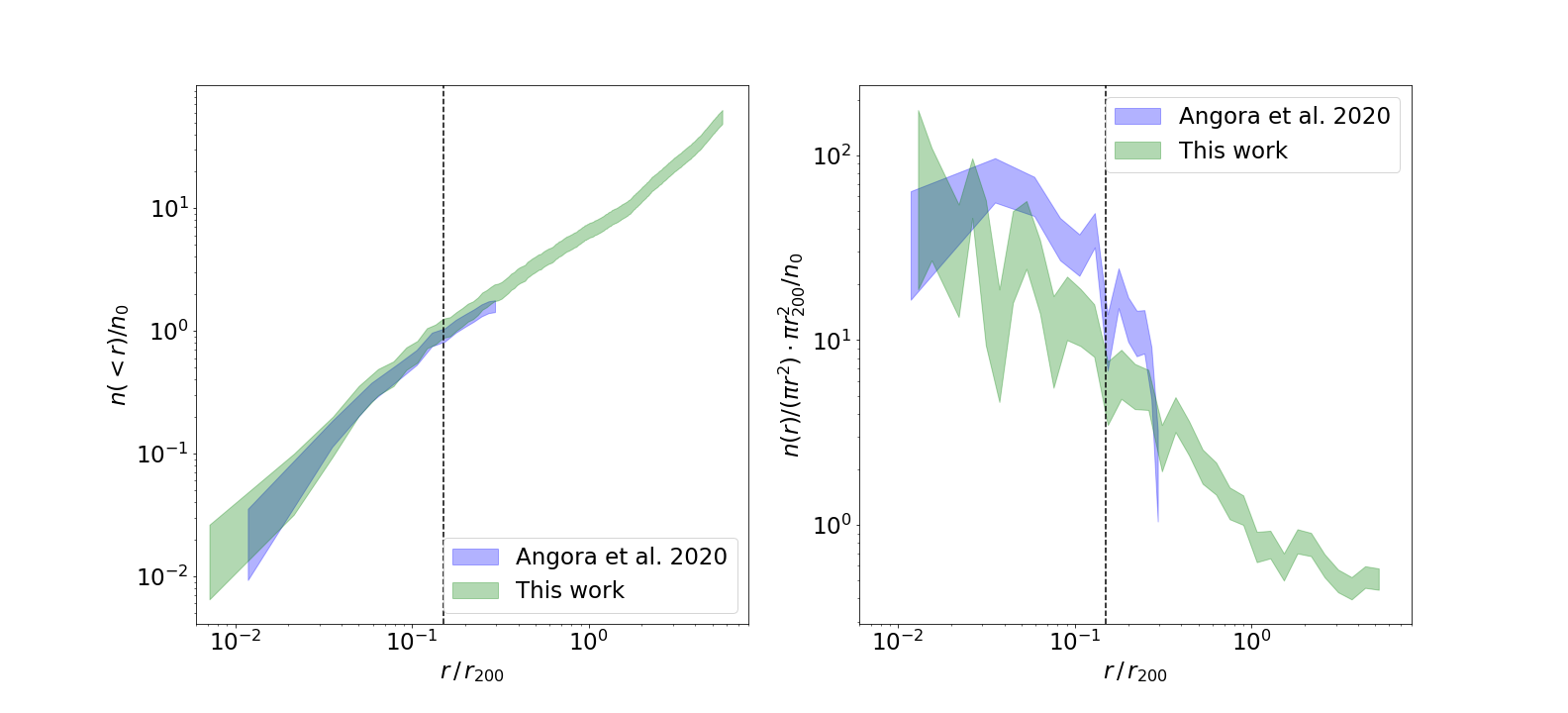}
\caption{Cumulative (\textit{left}) and differential (\textit{right}) projected number density of M0416 members identified in our cluster interval (green), compared with results obtained by \citet{Angora2020} (blue). The areas correspond to the 68\% confidence level regions. All profiles are normalized by the number density $n_0$ of members with r<0.15 r$_{200}$. The dashed line corresponds to r = 0.15 r$_{200}$.}
\label{fig:radial_histogram}
\end{figure*}

Additionally, we used our galaxy sample to estimate the cumulative projected number of cluster members and the differential number of density profiles. We computed the profiles as a function of the projected radius in units of r$_{200}$ and rescaled them by the number of members, $n_0$, found within the radius of r/r$_{200}=0.15$ to follow the convention implemented by \citet{Angora2020}. In Fig. \ref{fig:radial_histogram} we show the cumulative projected number and the differential projected number density profiles of cluster members after applying such a renormalization, where the shaded areas correspond to 68\% confidence levels. The slope of the cumulative profile reveals the presence of a structure, denser in the core, that is overall less dense but still structured up to the periphery of the field. The obtained profile follows the same distribution found by \citet{Angora2020} for the four clusters Abell S1063 (z=0.347), M0416, MACSJ1206.2-0847 (z=0.438), and MACS J1149+2223 (z=0.544), which also reproduce the results found by \citet{Bonamigo2018} and \citet{Caminha2019} using strong lensing modeling.

\subsection{Galaxy sample according to local environment}
\label{sec:localenv}

\begin{figure*}
\centering
\includegraphics[width=19cm]{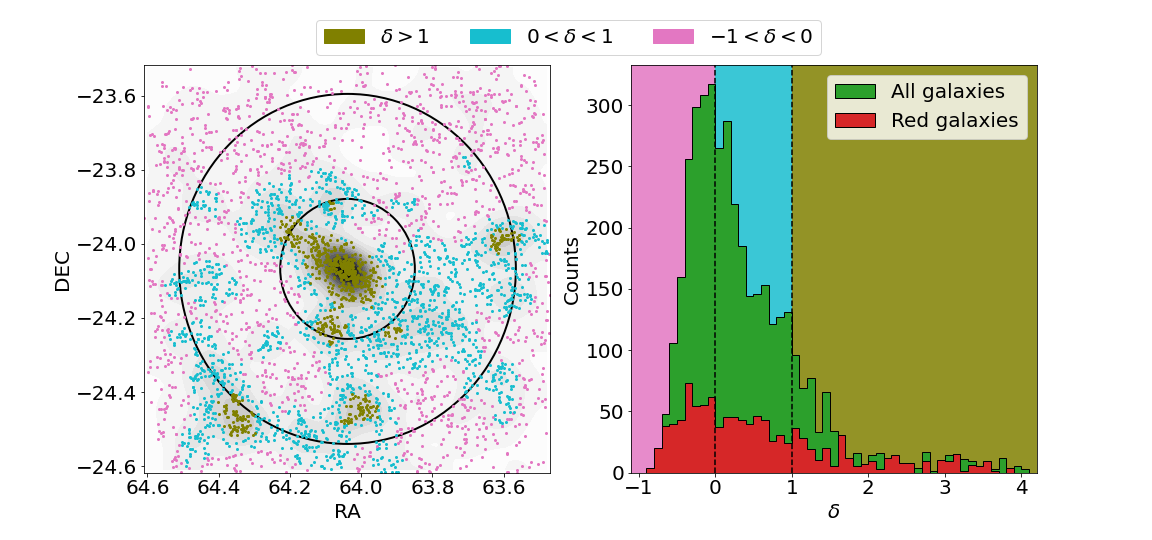}
\caption{Characterization of galaxies according to their local density. \textit{Left}: Environment classification of galaxies with $0.348<z_{phot}<0.453$ across the VST-GAME MACS0416 field. Colors correspond to the three density intervals chosen for this work. Black circles indicate 3.64 Mpc and 9.1 Mpc which correspond to 2r$_{200}$ and 5r$_{200}$, respectively. 
\textit{Right}: Distribution of galaxies according to their local density. The green histogram is computed with all the galaxies within the cluster interval, while the red histogram is made with red galaxies. Filled stripes identify the density intervals chosen as environment tracers.}
\label{fig:histodensity}
\end{figure*}

\begin{figure*}
\centering
\includegraphics[width=19.5cm]{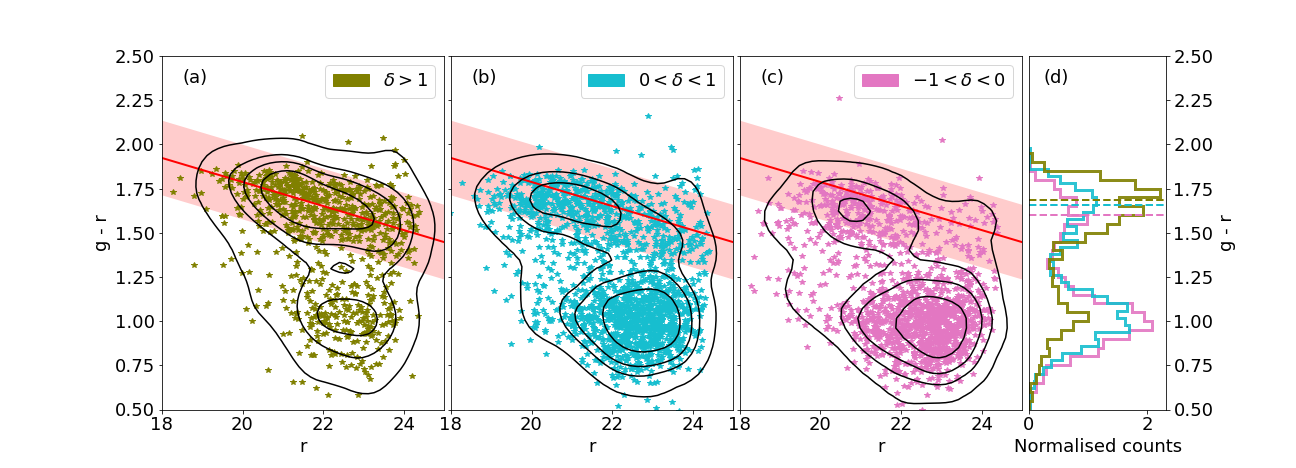}
\caption{Analysis of color-magnitude diagram dividing the galaxy sample (gray points of Fig. \ref{fig:colmag}) according to the local environment. The color-code follows that of Fig. \ref{fig:histodensity}. 
Panel (a) contains galaxies in the densest environments of the field, including the cluster core and outskirts overdensities. Panel (b) contains galaxies in denser environments up to two times the mean density. Panel (c) contains galaxies in environments less dense than the mean density of the field. Panel (d) shows a normalized histogram of the $g - r$ color for the three galaxy samples. Horizontal dashed lines represent the location of the peaks of the distributions (see text for details). The percentage of red galaxies is $65.2 \%$, $32.8 \%$, and $21.1 \%$ for the three density environments respectively, showing a clear abundance of red galaxies in the densest regions of the field. The black contours in each plot contain 5\%, 25\%, 50\%, and 75\% of the galaxies for each sample. The red line and the red area represent the sequence of red galaxies obtained with the spectroscopic cluster sample.}
\label{fig:rs_comparing_densities}
\end{figure*}

To characterize the local environment, we split the sample of all cluster members (regardless of color) into three density intervals: low ($\delta<0$), medium ($0<\delta<1$), and high ($\delta>1$). The first interval corresponds to underdense regions, with a local density (see Sec. \ref{sec:overdensity}) lower than the mean value over the field, $\bar{n}=0.92$; the second interval corresponds to overdense regions with a higher local density, up to two times the mean; and the third interval corresponds to regions with the highest density values over the field.
The first density bin contains 1507 galaxies (37\%), the second 1778 (44\%), and the third 782 (19\%).

The left panel of Fig. \ref{fig:histodensity} shows the spatial distribution of galaxies and highlights the different density intervals, while the right panel of the same figure shows the distribution of galaxies as a function of the density value.

To analyze the properties of galaxies in the different environments we compare the location of the red sequence determined in Sect. \ref{sec:rs} with the sample of galaxies in each of the three density intervals defined above. Figure \ref{fig:rs_comparing_densities} clearly shows the dependency of the color-magnitude diagram on the environment. The fraction of galaxies on (below) the red sequence increases (decreases) with density: the fraction of red galaxies goes from 21\% $\pm$ 1\% in the lowest density bin to  32 \% $\pm$ 1\% at intermediate densities to 65 \% $\pm$ 2\% in the highest density bin. Errors are binomial.

This behavior is also clearly visible in the normalized histograms (right panel of Fig. \ref{fig:rs_comparing_densities}) of the $g-r$ color. All of them present a bimodal behavior with one peak for the galaxies populating the red sequence and the other one the blue cloud, but the relative importance of the two peaks in the distribution strongly varies with the environment. The histogram also shows that the position of the red sequence slightly shifts with the environment: the peak of the distribution of red galaxies shifts toward redder colors with increasing density. Horizontal dashed lines in panel (d) of Fig. \ref{fig:rs_comparing_densities} show the position of the peak obtained by fitting a kernel density estimator (\textit{kde}) to the red galaxy histogram. 

\subsection{Galaxy sample on outskirts substructures}
\label{sec:outskirtstruct}

\begin{figure*}
\centering
\includegraphics[width=19cm]{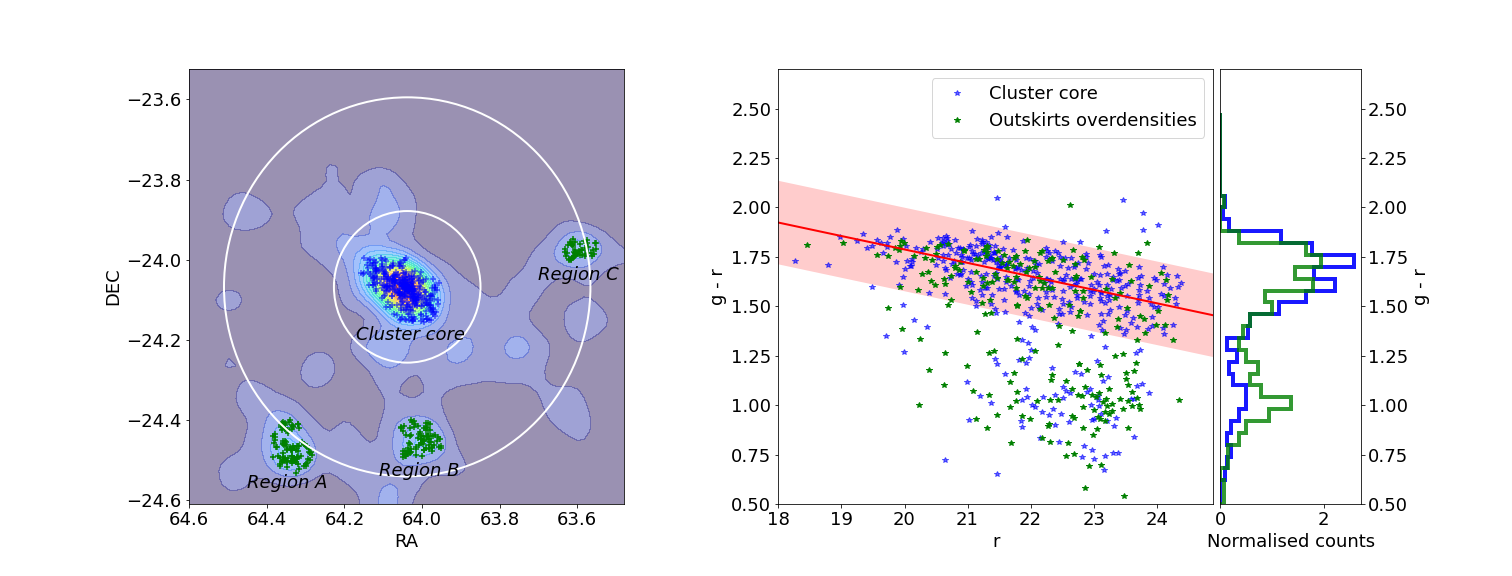}
\caption{Analysis of galaxy properties on outskirts overdensities. \textit{Left}: Selection of the three overdense regions in the cluster outskirts r>2r$_{200}$. The background shows the density field computed with the red galaxy sample (bottom panel of Fig. \ref{fig:radec_zphot}). The crosses over the plot correspond to galaxies in overdense environments with $\delta>1$. Regions A, B, and C are determined using the red-sequence density field. White circles are the same as those in Fig \ref{fig:radec_zphot}. 
\textit{Middle}: Detailed analysis of the color-magnitude diagram shown on panel (a) of Fig. \ref{fig:rs_comparing_densities}. The red stripe represents the red sequence found with spectroscopic cluster members. Blue points are galaxies with $\delta >1.5$, which corresponds to the cluster core. Green points are galaxies related to the three sub-structures (Regions A, B, and C) in the cluster outskirts at r>2r$_{200}$. 
\textit{Right}: Histogram of the $g-r$ color for both galaxy samples shown in the central panel.}
\label{fig:rs_core_vs_regions}
\end{figure*}

Thanks to the large FoV of the VST and VISTA images we can also investigate the outskirts of M0416 in detail. The left panel of Fig. \ref{fig:histodensity} unveiled the presence of three overdense regions in the cluster outskirts at r$\sim$5r$_{200}$ with more than 25 galaxies each. 
To better characterize their shape and galaxy content, we delimited the area of each region, defining a polygon that includes all the red-sequence galaxies with $\delta_{rs}>1$. Regions A, B, and C are shown in the left panel of Fig. \ref{fig:rs_core_vs_regions} and their properties are presented in Table \ref{table:regionsdata}.

\begin{table*}
\caption{Properties of the three overdense regions in the cluster outskirts.} 
\label{table:regionsdata}  
\centering                        
\begin{tabular}{c c c c c c}       
\toprule        
 Region & RA & DEC & Distance from core & Red galaxies & Blue galaxies \\
\midrule  
\rowcol Region A & 64:20:07.83 & -24:28:06.09 & 9.3 Mpc & 62 & 41 \\
Region B & 64:00:06.39 & -24:27:14.00 & 6.7 Mpc & 46 & 35 \\
\rowcol Region C & 63:35:56.84 & -23:58:30.23 & 9.1 Mpc & 27 & 21 \\
\hline
\end{tabular}
\tablefoot{RA and DEC coordinates indicate the geometrical center of the overdense region. Distance from the cluster center is also referred to as the geometrical center of each region.}
\end{table*}

The middle panel of Fig. \ref{fig:rs_core_vs_regions} shows the color-magnitude diagram for the galaxies inside the cluster core and outskirts' overdensities, and the right panel the corresponding $g-r$ color histogram is normalized. The most interesting feature of this color distribution is the presence of a similar fraction of red galaxies in the outskirts' overdensities to that in the cluster core. In the figure, the three outskirts' overdensities are shown together to increase the signal. Using a Kolmogorov-Smirnov test, we cannot reject the hypothesis that the color distribution of each of the three regions is generated from the same distribution. Nevertheless, an excess of blue galaxies is observed in region C (see Table \ref{table:regionsdata}). The luminosity distribution of the three outer regions has a median of 22.34, 21.85, and 22.30 mag in the $r$ band, respectively, showing that region B is slightly brighter. The mean luminosities of these three structures are similar to that of the cluster core, with a median luminosity of 22.07 $r$ mag in the same band.

\section{Discussion}
\label{sec:discussion}

As anticipated in the introduction, M0416 has already been largely analyzed in the literature; however, most of these studies have focused on the central core r$<$r$_{200}$ \citep{ebe01,pos12,Rosati14,ogr15,Merlin2016,bal16,castellano2016,Vulcani2016,Vulcani2017,Natarajan2017,lot17,annunziatella2017,Shipley2018} or were dedicated to characterizing the surface mass density from weak- and strong-lensing analyses \citep{zit13,jau15,Grillo15,bon17,Caminha17,Gonzalez20}. 

Only \citet{bal16} and \citet{OlaveRojas2018} examined the cluster outskirts. These studies identified spectroscopically confirmed substructures up to r $\sim$ 2 r$_{200}$, exploiting the \cite{DresslerShectman88} technique to isolate regions kinematically distinct from the main galaxy cluster \citep[see also ][]{Dressler13}.

The results presented in this paper show the importance of data covering regions up to 5r$_{200}$ to investigate galaxy evolution in different cluster environments. Only the grasp and image quality of the optical VST data combined with the NIR VISTA images, allowed us to carry out the investigation of galaxy properties in the full range of environments, from the high-density cluster core to the outskirts. The importance of cluster outskirts up to these distances lies in the fact that they allow an environmental study of infalling populations without the contamination from backsplash galaxies \citep{Jaffe2018,Haines15}. The $0.348<z_{phot}<0.453$ redshift interval is equivalent to a transverse distance of $\sim$ 360 Mpc, which marks the importance of a spectroscopic follow-up. 

In Sec. \ref{sec:overdensity} we present, for the first time, the large-scale density field for M0416, up to $\sim$5r$_{200}$, using deep photometry (mag $r$ < 24.4). Thanks to the density field shown in Fig. \ref{fig:radec_zphot}, we divided galaxies according to the local environment (see Sect.~\ref{sec:localenv}). We study the dependence of galaxy colors as a function of local densities to obtain a characterization of the properties of galaxy populations as a function of the environments. We show that, as expected, galaxies in less dense regions are mainly blue (Fig. \ref{fig:rs_comparing_densities}). We find that the peak of the distribution of red galaxies shows a shift toward redder colors with increasing density. This indicates the active role of the environment as a driver of galaxy evolution. The role of the masses in this process remains to be studied and represents one of our next steps. 

Additionally, in Sect.~\ref{sec:outskirtstruct}, we find the presence of three overdense regions (indicated as A, B, and C) at large distances ($\sim$ 5 r$_{200}$) from the cluster center and a large overdensity region aligned with the cluster core. We find that the three overdensities have mean densities and luminosities similar to the cluster core. Moreover, the galaxies populating these external overdensities are typically as red as galaxies in the cluster core (Fig. \ref{fig:rs_core_vs_regions}). We can speculate that these could be structures (e.g., dense groups or cluster cores) that are in the processes of being incorporated in the cluster through filaments. The color distribution also suggests the presence of evolved galaxy populations, an insight into a pre-processing phenomena over these substructures. In the next work, we will investigate complementary tracers for passive or star-forming galaxies, such as the color-color UVJ diagram, to better characterize the quenching scenarios on the cluster outskirts.

All the substructures identified by \citet{OlaveRojas2018} are placed in the overdense regions in our field ($0<\delta<1$ and $\delta>1$). The regions A, B, and C are outside the area covered in \citet{OlaveRojas2018}. The typical number of members in the substructures found in that work is nine, thanks to the dynamical analysis done with spectroscopic data. Overdense regions in this work typically contain $\sim$ 70 galaxies due to the choice of the local environment (cell size, kernel length) influenced by uncertainties on photometric redshifts. The fraction of red galaxies in substructures in \citet{OlaveRojas2018} is intermediate between that of the main cluster and the field, supporting the pre-processing picture. We find that the fraction of red galaxies on the overdensities is lower but similar to that of the cluster core, but higher than the field.

\citet{Schirmer11} obtained a density field up to 10 Mpc from the cluster SCL2243 at $z$=0.45 using both spectroscopic and photometric data and including filaments detection using a weak lensing analysis. In the filaments, they observe a constant color, independent of the clustercentric distance across the field. Only in the cluster infall region (out to 1.5 Mpc outside r$_{200}$) do the filaments become noticeably redder, having the same average color as the supercluster center. Instead, in this work, we claim overdensities with the same average color as the cluster center up to larger distances (10 Mpc $\sim$5r$_{200}$). This can be explained because the scale used to compute the density field in this work is spatially larger than the filament scale. 

\citet{Verdugo12} computed a density field for the cluster RXJ1347.5-1145 at $z=0.45$ on a scale of $\sim20 \times 20$ Mpc. They also found dense regions with a low fraction of blue galaxies up to $\sim$ 10 Mpc from the cluster core in perfect agreement with our work. Still, their optical photometry is shallower than ours  ($r$ mag = 23.5 vs mag = 24.4), which allows us to explore a wider range of masses. To exploit the depth of our data, in a forthcoming paper we will compute the stellar masses of our sample and study their dependence on different galaxy properties. 

\citet{lu12} studied $\sim$100 galaxy clusters between $0.16<z<0.36$ finding that the fraction of optically blue galaxies is lower for the overdense galaxy population in the cluster outskirts compared to the average field value, at all stellar masses $M_{*} > 10^{9.8} M_{\odot}$ and at all radii out to at least 7 Mpc, which is in agreement with our results. Nevertheless, with our data, we reach larger distances ($\sim$10 Mpc at the same redshift). In future work, we will compute stellar masses, and we expect to extend this result to less massive galaxies, $M_{*} \sim 10^{9} M_{\odot}$ since our photometry is deeper by 0.5 mag in the $r$ band. 

\citet{Just2019} studied 21 galaxy clusters using spectroscopic data at $0.4<z<0.8$ up to $r$ mag = 22.9, finding that galaxies in the infall regions, determined dynamically, show enhanced clustering. They also found that the more highly clustered galaxies show an elevated red fraction, which is interpreted as pre-processing. Our analysis does not determine the infalling regions through spectroscopic data. Nevertheless, we also find that highly clustered galaxies, up to large radial distances, show an elevated red fraction.

\citet{Sarron19} claimed to detect cosmic filaments around galaxy clusters using photometric redshifts in the range $0.15 < z < 0.7$ and found that the fraction of passive galaxies is higher in filaments than in isotropically selected regions around clusters and that the passive fraction in filaments decreases with increasing distance to the cluster up to $\sim$ 5 Mpc. Our present work does not focus on filaments, but current spectroscopic observations (see below) of our M0416 field will allow us to robustly identify filaments and to study their galaxy content up to $\sim$ 10 Mpc from the cluster core.
\\
Here, we report various observational hints that have unveiled the unmistakable role of the environment in accelerating galaxy evolution at intermediate redshift. Indeed, several works have found that dense environments are populated with red galaxies, from the local universe up to $z\sim1.5$ \citep[e.g.][]{Haines2007,Cooper2010,Pasquali2010,Peng2010,Peng2012,McGee2011,Sobral2011,Muzzin2012,Smith2012,Wetzel2012,LaBarbera2014,Lin2014,Vulcani2015,vanderburg2020}, and our results strengthen the role of the large-scale structure in aging galaxies across cosmic times.

\section{Summary}
\label{sec:summary}

This paper was written in the context of the VST-GAME survey, a project aimed at gathering observations at optical (VST) data for six massive galaxy clusters at 0.2$\lesssim$z$\lesssim$0.6. The goal is to investigate galaxy evolution as a function of stellar mass and environment. 

In this work, we focused on M0416, presenting the procedure adopted to extract the multiband photometry and release the catalog. The same approach will be then applied to all the other clusters of the survey and the corresponding catalogs will be presented elsewhere.

With this paper, we make the catalog presented in Sect. \ref{sec:multiband} containing the main photometric parameters of the survey available to the community. This catalog is the result of a prudent process of source extraction and band matching, which guarantees its quality, penalizing the total number of objects: $74114$ in total. Single-band catalogs, available upon request, contain a larger number of objects, such as $237094$ in the $r$-band, but over the completeness limit or without a clear counterpart in the other bands. 

We applied \textit{LePhare} to the multiband catalog to derive photometric redshifts for all galaxies. Spectroscopic redshifts in the central part of the cluster were used to calibrate the SED fitting procedure and assess the quality of the obtained photometric redshifts.

Overall, we obtained the following statistics to quantify the accuracy of the photo-$z$'s: $bias = 0.0364$, $\sigma = 0.0296$, $\sigma_{NMAD} = 0.0425$, $\eta = 3.75\%$. Galaxies with an S/N>5 were assigned to the cluster if their photo-$z$ is within $0.348 < z_{phot} < 0.453$.
We estimated the completeness and purity of the photometric sample on that interval, with respect to the whole spectroscopic sample, finding 63.3 \% and  55.8 \%, respectively. Considering only red galaxies, the completeness and the purity increase to 100.0\% and 71.1\%,  respectively, concerning the red spectroscopic sample.

In Sect.~\ref{sec:Cluster_membership}, we show the cluster membership and discuss the reliability of this selection. As additional validation of the cluster membership selection, we analyzed the radial distribution of galaxy density for M0416 (Fig. \ref{fig:radial_histogram}). The obtained profile follows the same distribution found by \citet{Angora2020} using four CLASH-VLT clusters, including M0416.
This confirms that our density field is dominated by the cluster core at the center of the field, but it also supports the presence of structure up to 5 r$_{200}$. 

In the second part of the paper, we give an environmental analysis to identify regions of different densities.

We unveil the presence of overdense regions, also at large distances ($\sim$5r$_{200}$) from the cluster center, and a large overdensity region aligned with the cluster core. By comparing the location of the red sequence determined in Sect. \ref{sec:rs} with the sample of galaxies in different environments, we find that there is a dependency of the color-magnitude diagram on the environment. The fraction of galaxies on (below) the red sequence increases (decreases) with density, the fraction of red galaxies goes from 21\% $\pm$ 1\% in the lowest density bin to  32 \% $\pm$ 1\% at intermediate densities, to 65 \% $\pm$ 2\% in the highest density bin, considering binomial errors.
This behavior is also clearly visible in the normalized histograms in panel (d) of Fig.~\ref{fig:rs_comparing_densities}, where the relative importance of the two peaks at red and blue colors strongly varies with the environment. The histogram also shows that the peak of the distribution at the reddest colors overall shifts toward redder colors with increasing density.
\\
The procedure developed here will be applied to the other five clusters of the VST-GAME survey. The results of a spectroscopic follow-up for the M0416 outskirts are under preparation (P.I. A. Mercurio, with 2dF + AAOmega, semester 2021B). These new data will enable the study of spectral properties (i.e. only absorption or also the presence of emission lines) and their association with the cluster environments: substructures, and infalling filaments. 

In light of our results, and considering them in the open scenario of galaxy evolution that is still under construction, we can conclude that we exceed the initial aims of this paper. We proved the presence of environmental effects on galaxy colors in intermediate-redshift cluster outskirts at distances almost unexplored until now.  This opens a range of possibilities for further studies, with the unique VST-GAME data or with new-generation instruments. 

\bibliographystyle{aa}
\bibliography{bib_game}

\begin{acknowledgements}
We are grateful to the anonymous referee for the suggestions and comments that helped to improve this paper.
N.E. acknowledge F. Santoliquido, F. Sinigaglia, R.S. Remus, and P. Jablonka for precious talks during the development of this work. N.E. acknowledge the support from the Cariparo foundation. We acknowledge financial contributions by PRIN-MIUR 2017WSCC32 "Zooming into dark matter and proto-galaxies with massive lensing clusters" (P.I.: P.Rosati), INAF ``main-stream'' 1.05.01.86.20: "Deep and wide view of galaxy clusters (P.I.: M. Nonino)" and INAF ``main-stream'' 1.05.01.86.31 "The deepest view of high-redshift galaxies and globular cluster precursors in the early Universe" (P.I.: E. Vanzella). G.R. and B.V. acknowledge the support from grant PRIN MIUR 2017 - 20173ML3WW 001. R.D. gratefully acknowledges support by the ANID BASAL projects ACE210002 and FB210003. B.C.L. acknowledges support from the National Science Foundation under Grant No. 1908422.
\end{acknowledgements}

\newpage

\begin{appendix}

\section{Masking procedure}
\label{appendix:a}

In this appendix, we explain the procedure used to mask haloes and ghosts across the VST and VISTA fields in detail. The coefficients presented in this section are calibrated for each band and can be used in photometric surveys with a similar deepness. 

Haloes are always centered on the star that generates them, and their radius, r$_{halo}$ in pixels, linearly depends on the magnitude of the star; r$_{halo}=b_1 \times mag + b_2$, where $b_1=-150.29$ and $b_2=1821.34$, according to the handmade haloes drawn for the $r$ band. Table \ref{table:coef_haloes_b} presents, for all the bands, the coefficients to determine the halo radius ($b_1$ and $b_2$).

The location of the ghosts depends on the position of the parent star in the OmegaCAM field. They are aligned in the radial direction, towards the external regions, from the center of the field with the generating star. The distance from the star to the center of the ghost ($d_{SG}$ in pixels) depends linearly on the distance between the center of the field and the star ($d_{CS}$): $d_{SG}=c_1 \times d_{CS} +c_2$, where $d_{center-star}$ is measured in pixels, and the coefficients for the $r$ band are $c_1=0.07716$ and $c_2=38.16$. The radius of the ghost depends on the GAIA $G$ magnitude of the star: 1000 pixels for $G\le8.7$, 900 pixels for $8.7 < G\le9.7$, 850 pixels for $9.7 < G\le10.2$, and 200 pixels for $10.2 < G\le13$. Following this procedure, we find that for stars near the center of the field, haloes and ghosts are usually overlapped, while for stars in the outskirts, the two spurious regions are separate. 

The size of haloes and ghosts depends on the band, being larger in the $r$, $g$, and $i$ bands. The $u$ band presents the smallest spurious regions of the VST OmegaCAM fields. On the other hand, VISTA fields with VIRCAM present similar behavior, but the haloes and ghosts are smaller. Table \ref{table:coef_haloes_c} presents the coefficients to determine the ghost positions ($c_1$ and $c_2$) and the ghost size as a function of the magnitude of the star in the GAIA $G$ band for all the bands.

\begin{table}[h]
\caption{Coefficients used to determine the radius of the halo.} 
\label{table:coef_haloes_b}  
\centering                        
\begin{tabular}{c c c}       
\toprule
Band & $b_1$ & $b_2$ \\
\midrule
\rowcol$u$ & -108.78 & 1338.92 \\
$g$ & -130.76 & 1573.16 \\
\rowcol$r$ & -150.29 & 1821.34 \\
$i$ & -116.58 & 1389.50 \\
\\
\rowcol$Y$  & -50.59 & 694.38 \\
$J$  & -71.82 & 920.87 \\
\rowcol$Ks$ & -48.12 & 648.73 \\
\hline
\end{tabular}
\tablefoot{The radius of the halo is measured in pixels. Haloes are generated by stars brighter than 13 mag in the GAIA $G$ band}
\end{table}

\begin{table}[h]
\caption{Coefficients used to mask ghosts.} 
\label{table:coef_haloes_c}  
\centering                        
\begin{tabular}{c c c c }       
\toprule
Band & $c_1$ & $c_2$ & ghost radius \\
\midrule
\rowcol$u$ & 0.09295 & -89.92 & 900, 400, 400, 200\\
$g$ & 0.07729 & 29.27 & 1000, 900, 850, 200\\
\rowcol$r$ & 0.07716 & 38.16 & 1000, 900, 850, 200\\
$i$ & 0.07145 & 79.50 & 950, 900, 850, 200\\
\\
\rowcol$Y$  & 0.01651 & 38.70 & 400, 400, 400, 300\\
$J$  & 0.00459 & 14.88 & 600, 500, 400, 300\\
\rowcol$Ks$ & 0.00470 & -2.57 & 400, 300, 300, 300\\
\hline
\end{tabular}
\tablefoot{The ghost radius, and the distance from the generating star to the ghost are measured in pixels. The radius of the ghost is related to the GAIA $G$-band magnitude in the following intervals: $G\le8.7$, $8.7 < G\le9.7$, $9.7 < G\le10.2$, and $10.2 < G\le13$.}
\end{table}
\end{appendix}

%
%

\end{document}